\documentclass[nofootinbib,superscriptaddress,
floatfix,twocolumn,aps,pra,10pt]{revtex4-1}

\usepackage[utf8]{inputenc}
\usepackage[T1]{fontenc}

\usepackage{amsmath,amsfonts,amssymb}
\usepackage{bbm}
\usepackage{color}
\usepackage{textcomp}
\usepackage[normalem]{ulem}
\usepackage{pifont}
\usepackage{multirow}
\usepackage{graphicx}
\usepackage{enumerate}	

\usepackage{hyperref}

\newcommand{\be}{\begin{equation}}
\newcommand{\ee}{\end{equation}}
\newcommand{\ra}{\rangle}
\newcommand{\la}{\langle}
\newcommand{\ket}[1]{\lvert#1\ra}
\newcommand{\bra}[1]{\la#1\rvert}
\newcommand{\braket}[2]{\la#1|#2\ra}
\newcommand{\proj}[1]{\ket{#1}\!\bra{#1}}

\newcommand{\abs}[1]{\lvert#1\rvert}

\newcommand{\nn}{\nonumber}

\begin{document}

\title{Quantum walks on embedded hypercubes: Nonsymmetric and nonlocal cases}

\author{Adi Makmal}
\email{adi.makmal@uibk.ac.at}
\affiliation{Institut für Theoretische Physik,
Universität Innsbruck, Technikerstr.\ 21A, A-6020 Innsbruck, Austria}
\affiliation{Institut für Quantenoptik und Quanteninformation,
Österreichische Akademie der Wissenschaften, Technikerstr.\ 21A, A-6020 Innsbruck, Austria}

\author{Markus Tiersch}
\affiliation{Institut für Theoretische Physik,
Universität Innsbruck, Technikerstr.\ 21A, A-6020 Innsbruck, Austria}

\author{Clemens Ganahl}
\affiliation{Institut für Theoretische Physik,
Universität Innsbruck, Technikerstr.\ 21A, A-6020 Innsbruck, Austria}

\author{Hans J. Briegel}
\email{hans.briegel@uibk.ac.at}
\affiliation{Institut für Theoretische Physik,
Universität Innsbruck, Technikerstr.\ 21A, A-6020 Innsbruck, Austria}

\date{\today}

\begin{abstract}
The expected hitting time of discrete quantum walks on a hypercube (HC) is numerically known to be exponentially shorter than that of their classical analogs in terms of the scaling with the HC dimension. Recent numerical analyses illustrated that this scaling exists not only on the bare HC, but also when the HC graph is symmetrically and locally embedded into larger graphs.
The present work investigates the necessity of symmetry and locality for the speed-up by considering embeddings that are nonsymmetric or nonlocal. We provide numerical evidence that the exponential speed-up survives also in these cases. Furthermore, our numerical simulations demonstrate that removing a single edge from the HC also does not destroy the exponential speed-up. In the nonlocal embedding of the HC we encounter dark states, which we analyze.
We provide a general and detailed presentation of the mapping that reduces the exponentially large Hilbert space of the quantum walk to an effective subspace of polynomial scaling. This mapping is our essential tool to numerically study quantum walks in such high-dimensional structures.
\end{abstract}

\maketitle

\section{Introduction} \label{sec:intro}

Quantum walks~\cite{2003_Kempe_review, 2012_Reitzner_review} on hypercube (HC) graphs and on glued trees exhibit hitting times that scale exponentially faster than the classical analogs~\cite{2003_Kempe, 2003_Childs}.
For discrete quantum walks \cite{2000_Nayak_Arx,2001_Ambainis_Proceedings, 2001_Aharonov_Proceedings} it was numerically shown that the expected hitting time also exhibits an exponential speed-up~\cite{2006a_Krovi_PRA} even for a slightly distorted HC.
In a recent study~\cite{2014_HCI_PRA} it was numerically shown that such an exponential speed-up for the expected hitting time on the HC can also be obtained for large families of HC graphs that were embedded in larger graphs, including HCs where additional tails (linear graphs) were appended to every HC vertex and HCs that were appended (recursively) with HCs. These two families of structures have two common features: (a) They retain the permutational symmetry of the HC, i.e., all vertices of the central HC have the same structure attached, and (b) the embedding is local, meaning that the external structures are connected to the HC only via a single vertex, hence there is no direct connection between these attached structures other than the HC.

The exponential speed-up of the quantum walk originates from constructive and destructive interference patterns that result in the particle appearing at the target vertex on the opposite side of the HC with a large probability. The specific structure of the graph, in particular its inherent symmetry or lack thereof, also plays a crucial role in forming the interference pattern~\cite{2006b_Krovi_PRA}. It is thus of interest to investigate to what extent the exponential speed-up of the quantum walk on the embedded HC is maintained when the embedding is not symmetric. Similarly, it is of interest to clarify whether the embedding must be local to maintain the speed-up.  

In this paper we study the necessity of symmetry and locality in the embedding procedure for the sake of obtaining an exponential speed-up. To that end, we consider a nonsymmetric embedding of the HC in the form of a HC with a single vertex attached to it, and a nonlocal embedding in the form of a HC that is embedded in another HC of a higher dimension. We also analyze the case where a single edge is removed from the HC. For all three setups we analyze the scaling of the expected hitting times for the classical and the discrete quantum walks with analytical and numerical tools, respectively.

For HCs of high dimension $d$, the numerical analysis of the expected hitting time (of both the classical and the quantum walks) becomes intractable because the number of vertices in the graph grows exponentially with $d$. In order to facilitate such calculations, it is expedient to map the walk on the full HC graph to an equivalent walk on a smaller structure by using the symmetry of the HC. For the case of the bare HC, such a procedure results in a walk on a line of $d+1$ vertices~\cite{Kac}. However, when the HC is perturbed in a nonsymmetric way, this known mapping to the line no longer holds. We therefore generalize this procedure and show that a walk on a HC for which $m$ vertices are modified can be mapped to a walk on an $(m+1)$-dimensional grid. The size of the grid in each dimension is at most $d+1$. A constant number of modifications thus leads to a finite grid with a number of vertices that grows only polynomially with the HC dimension~$d$. 
For the quantum walk, this procedure includes a mapping of the full Hilbert space to an effective subspace, on which the walk takes place (see \cite{2002_Moore_inBook, 2006a_Krovi_PRA} for such a mapping of the quantum walk on the bare HC to a line).

The paper is structured as follows.
Section~\ref{sec:preliminaries} presents the necessary formalism and definitions for classical and quantum walks.
In Sec.~\ref{sec:mapping} we introduce a general mapping for walks on HCs with several perturbed vertices to walks on lower-dimensional grids. We further show how to map the Hilbert space of the quantum walk to a lower-dimensional subspace. 
In Sec.~\ref{sec:results} we present the numerical results for various scenarios of nonsymmetric and nonlocal extensions of the HC graph.
Section~\ref{sec:conclusion} summarizes the paper.

\section{Preliminaries} \label{sec:preliminaries}
\subsection{Classical random walk on the embedded HC} \label{secsec:classical_prliminaries}
The classical expected hitting time of a discrete random walk on a graph from vertex $x_0$ to vertex $x_f$ is defined as 
\be \label{eq:ht_infinite_sum}
	\tau_{x_0x_f}\equiv\tau(x_0) = \sum_{t=0}^{\infty} t\; p_{x_f}(t), 
\ee
where $p_{x_f}(t)$ is the probability to hit the final vertex $x_f$ for the first time at time step $t$.
A random walk on the graph with a target at $x_f$ is equivalent to an absorbing Markov chain~\cite{GrinsteadSnell1997}, which has probability 1 to remain on this final vertex. Its transition matrix $[P]_{ij} = p(x_i \to x_j)$ collects the probabilities to go from vertex $x_i$ to $x_j$ in one time step via all possible paths. The submatrix $Q$ of $P$ without the transitions to and from the final vertex gives rise to its fundamental matrix $F=(I-Q)^{-1}=\sum_{k=0}^\infty Q^k$, the $i$th row of which, when summed up, is one way to calculate the expected hitting time to reach $x_f$ when starting at vertex $x_i$.

When referring to vertices of a $d$-dimensional HC we will employ a notation that labels all vertices by a bit string of length $d$, i.e., $00\ldots0_2$ is the label of the starting vertex $x_0$ and $\overline{x_0} = 11\ldots1_2$ is its opposite corner. An important tool will be to group vertices according to the number of 1-symbols in their bit string, that is, their Hamming weight, which we will denote by $\abs{x}$ and which ranges from 0 for the starting vertex to $d$ for its opposite corner. The number of 1-symbols in which two vertices differ defines their Hamming distance.
Due to symmetry on the HC it is often sufficient to refer to its vertices only by their Hamming weight and we use the notation $\tau(x_0)\equiv\tau(0)$.

For the bare HC, the expected hitting time from one corner of the HC to the opposite corner is given by (see, e.g., \cite{2006a_Krovi_PRA})
\be \label{eq:classical_ht_hc}
	\tau(0) = \sum_{k=0}^{d-1}\Delta(k)= \sum_{k=0}^{d-1}  \frac{\sum_{j=0}^{k} {d\choose k-j}} {{d-1\choose k}},
\ee
where $\Delta(k)=\tau(k)-\tau(k+1)$, with the boundary conditions $\tau(d)=0$ and $\tau(d-1)=T_d-1=2^d-1$, in which $T_d$ denotes the so-called return time from any node to itself on a HC of dimension $d$. 
We will use this expression to obtain exact expected hitting times and bounds thereof for the classical walk. In particular, note that for the classical walk $\tau(0)>\tau(d-1)$,\footnote{Although the relation becomes intuitively clear when considering that for $\tau(0)$ the walker has to travel farther, the expression for $\tau(d-q)$ is explicitly given in Sec.~\ref{secsec:non_local}.} which is the origin of the exponential scaling of the expected hitting time of the classical walk on the bare HC, as well as on the structures we examine in this paper.

\subsection{Quantum walk} \label{secsec:quantum_prliminaries}

In what follows we consider the discrete coined quantum walk \cite{2000_Nayak_Arx, 2001_Ambainis_Proceedings, 2001_Aharonov_Proceedings} on undirected graphs $G=(V,E)$, with $V$ and $E$ being the set of vertices and edges, respectively. 
The Hilbert space is composed of a position space $\mathcal{H}_{P}$, spanned by the respective position states for all vertices $x_i \in V$ in the graph, and a coin space $\mathcal{H}_{C}$, spanned by the possible transition directions for each vertex:
\be
\label{eq:hc_hilbert_space}
\mathcal{H} = \bigoplus_{x_i \in V}\mathcal{H}_{C}^{x_i} \otimes \mathcal{H}_{P}^{x_i},
\ee
where $\mathcal{H}_{P}^{x_i}$ is the position Hilbert space for vertex $x_i$ and is spanned by the vector $\ket{x_i}$.
When the graph is regular, i.e., each vertex has the same number of neighbors, the coin spaces for all vertices are identical ($\mathcal{H}_{C}^{x_i}=\mathcal{H}_{C}$) and this expression reduces to a simple tensor product of the coin and the position space:
\be
\label{eq:regular_hc_hilbert_space_total}
\mathcal{H} = \mathcal{H}_{C} \otimes \mathcal{H}_{P}.
\ee
In this work, however, not all graphs are regular.

The walk unitary $U = SC$ is the composition of a shift and a coin operator.
The shift operator is given by
\be \label{eq:shift_operator}
	S = \sum_{x_i\in V}\sum_{j=1}^{p_i}\ket{j,g(j,x_i)}\!\bra{j,x_i},
\ee
where $p_i$ is the degree of vertex $x_i$ and 
$g(j,x_i)$ gives the $j$th neighbor of vertex $x_i$ according to the graph connectivity.
The coin operator is given by
\be \label{eq:conditioned_coin_operator}
	C = \bigoplus_{x_i \in V} {C}_{p_i} \otimes \ket{x_i}\!\bra{x_i}, 
\ee
with
\be
\label{eq:coin_operator}
	C_{p_i} = \frac{2}{p_i} \!\!
\left(\!\! \begin{array}{cccc}
 1-\frac{p_i}{2} & 1 & \cdots& 1 \\
 1 &1-\frac{p_i}{2} & \cdots & 1 \\
 \vdots & \ddots & \ddots & 1 \\
 1 & 1 & \cdots & 1-\frac{p_i}{2}
 \end{array}\!\! \right),
\ee
where $C_{p_i}$ is the so-called Grover coin. If the graph is regular with degree $p$, the coin operator is reduced to $C=C_{p} \otimes I$.

Initially, the walker is placed on a starting vertex $x_0$ with a uniform probability to walk in every possible direction, resulting in the initial state
\be 
\ket{\psi_0} = \frac{1}{\sqrt{p_0}}\sum_{i=1}^{p_0}\ket{i} \otimes\ket{x_0}.
\ee
The target state is defined by the final vertex $x_f$ and assumes any arbitrary coin state, so there may be several final states $\{\ket{\psi_{f_1}}, \ket{\psi_{f_2}},\dotsc\}$. Yet not any coin state can be reached, and for walks on the HC, the symmetry of the walk unitary and the initial state dictates 
through which coin states the walker can arrive at the final vertex. 
For example, if the final vertex is at the opposite corner of the HC, symmetry ensures that the final state is of the form \cite{2002_Moore_inBook}
\be 
\ket{\psi_f} = \frac{1}{\sqrt{p_0}}\sum_{i=1}^{p_0}\ket{i} \otimes\ket{x_f},
\ee
i.e., with a completely symmetric coin state, just like the initial state.

The hitting time of the quantum walk can be defined in several ways (see, e.g., \cite{2003_Kempe}). In the present work we employ the expected hitting time as defined in \eqref{eq:ht_infinite_sum} and used also in \cite{2006a_Krovi_PRA} (see \cite{2014_HCI_PRA} for a short discussion on the difference between different kinds of hitting time definitions). 
For the quantum walk the term $p_{x_f}(t)$ also denotes the probability to hit the target state $x_f$ for the first time at time $t$ and is given by
\be
	p_{x_f}(t) = \sum_j \abs{\bra{\psi_{f_j}}(U\Pi_0)^t\ket{\psi_0}}^2,
\ee
where $\ket{\psi_0}$ and $\ket{\psi_{f_j}}$ are the initial and final states of the walk (including both the positional and directional parts), respectively, $U$ is the walk unitary as defined before, and $\Pi_0 = I-\Pi_f$ is a projection operator, with $\Pi_f = I \otimes \proj{x_f}$, that projects to the final vertex $x_f$ for all coin states. Here the sum runs over all possible final states $\psi_{f_j}$ that share the same final vertex in the position part. These dynamics represent what is known as the measured quantum walk \cite{2003_Kempe}, where, after each application of the walk unitary, a partial measurement is preformed to check if the particle is found on the final vertex. If it is found, the walk is over; otherwise, the state of the particle is projected to the Hilbert subspace spanned by all but the final vertex.

While for graphs with a position space of low dimension the expected hitting time of the quantum walk can be calculated exactly, as shown in \cite{2006a_Krovi_PRA}, graphs with a Hilbert space of high dimension result in expressions that are numerically intractable. Therefore, we approximate the expected hitting time given as an infinite sum in \eqref{eq:ht_infinite_sum} with a finite sum
\be
	\label{eq:ht_finite_sum}
	\tau_{p_\text{tot}} := \sum_{t=0}^{T} t\, p_{x_f}(t),
\ee  
with summands until time step $T$, when a total of
\be
	\label{eq:total_prob}
	p_\text{tot}(T) = \sum_{t=0}^{T} p_{x_f}(t)
\ee
of the walker has arrived at the final vertex. 
Equation~\eqref{eq:ht_finite_sum} serves as a good approximation to the correct expected hitting time only if $p_\text{tot}=(1-\epsilon) \rightarrow 1$, that is, when $\epsilon \rightarrow 0$, and the remaining summands $tp_{x_f}(t)$ are small and decay fast enough for $t>T$.
In our numerics, we set $\epsilon$ such that the approximated expected hitting time $\tau_{p_\text{tot}}$ is well converged. That is, we run simulations with double precision and decrease $\epsilon$ by a factor of 10 until we no longer observe visual changes in our plots.
Throughout the paper different scenarios will entail different threshold values of~$\epsilon$.

\section{Reducing the quantum walk to an effective subspace} \label{sec:mapping}
\subsection{Mapping of the graph}

We consider several cases of modifying the HC: (i) by appending a graph nonsymmetrically to it, (ii) by embedding the HC nonlocally as a subgraph into a HC of higher dimension, or (iii) by deleting a single edge. All of these modifications destroy the original symmetry of the HC, but some symmetry is left, which we use to map the modified HC to a smaller graph, that is, in effect we identify the subspace in which the walk takes place. This procedure of mapping the walk to a lower-dimensional structure is the essential and important tool that we use for the numerical study of walks on modified HCs of high dimension (see~\cite{2014_Novo_arXiv} for a general computational approach to reducing the effective dimension of a quantum walk without prior knowledge of the symmetries of the graph).

The mapping is done in a way similar to the mapping of the unperturbed HC to a line.
The unperturbed HC graph is symmetric under permutation of the order of the bits in the bit string. This allows for a mapping of the corner-to-corner walk on the HC to a line, where the coordinate along the line represents the Hamming distance from the starting point (as well as the Hamming distance from the target vertex). This mapping corresponds to the Ehrenfest model from classical thermodynamics~\cite{Kac,Voit}. There $d$ gas particles can be located in either of two containers labeled $0$ or $1$ and initially all particles are located in the $0$-labeled container. At each time step a single gas particle is chosen uniformly at random and placed in the other container. The process ends when all particles are found at the $1$-labeled container. This process generates the corner-to-corner random walk on the $d$-dimensional HC, where the bit string indicates the location of the particles. The sum of particles in the $1$-labeled container then gives the random walk on the line with values $0$ to $d$.
In what follows we will treat perturbed HCs, which will generally result not in a mapping to a line but to a higher-dimensional grid. The random walk on such grids is in spirit similar to extensions of the Ehrenfest model to more types of particles (the Bernoulli-Laplace model; see, e.g., \cite{Donnelly}), where each direction in the grid counts the number of particles of a given kind in the $1$-labeled container, or higher-dimensional discrete-time variants of so-called linear birth-and-death processes (see, e.g., \cite{Hutton}). However, here we obtain transition probabilities at the perturbed vertices that are different from the above models.

\begin{figure}[t]
  \includegraphics{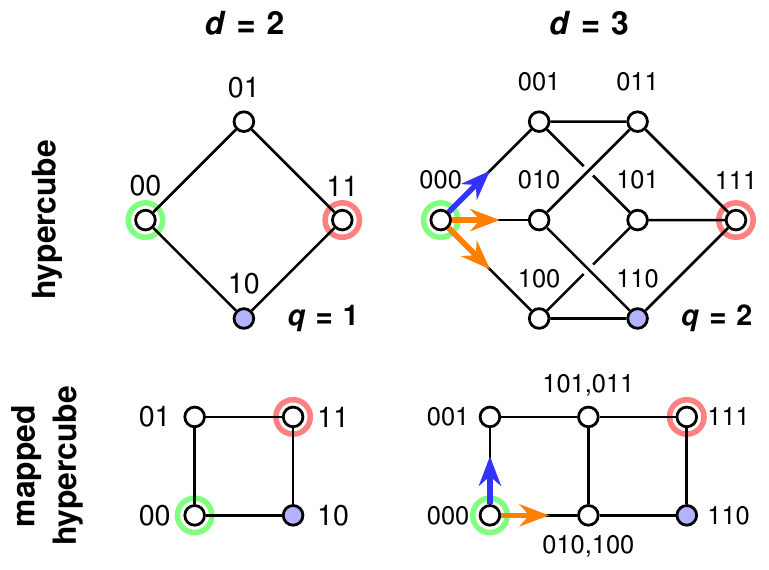}
  \caption{Mapping of HCs (dimension $d=2,3$) with a single modified vertex ($m=1$) to two-dimensional grids. Start and end vertices of the walk on the HC are left-most (Hamming weight 0) and right-most (Hamming weight $d$) vertices, circled in green and red, respectively. The perturbed vertex is filled in blue. For $d=3$ all three walking directions from the starting vertex are indicated by arrows: walking directions on the HC are mapped to horizontal and vertical directions on the grid.}
  \label{fig:mapping_illustration}
\end{figure}

The general procedure of mapping a full HC with modifications on $m$ vertices onto a smaller equivalent structure works as follows. First, all $m$ perturbed vertices are identified and marked according to their relative Hamming weights. This results in $m+1$ special vertices in the HC: the starting vertex $x_0=0$ and each of the $m$ modified vertices. These special vertices need to be treated individually and thus remain single vertices also in the mapped graph.
By symmetry, the bitwise complement of the modified vertices must be treated individually as well and therefore also remain single vertices in the mapped graph. 
The remaining vertices can then be grouped according to their connectivity and their Hamming distance from these special vertices. In general, each additional perturbation on the HC will increase the resulting dimension of the mapped graph by 1, so the reduced graph will be a grid of dimension $m+1$. For example, the bare HC (with no perturbations $m=0$) can be mapped to a one-dimensional line, whereas a HC with a single perturbation ($m=1$) can be mapped to a two-dimensional (2D) grid, as we show next.

In the scenarios we consider in this paper the number of modified vertices on the HC is given either by $m=1$ (for nonsymmetric and nonlocal embeddings, see Secs.~\ref{secsec:non_sym_emb} and \ref{secsec:non_local}, respectively) or by $m=2$ (for the deleted-edge scenario, see Sec.~\ref{secsec:removed_edge}). Here we explain the mapping in detail for the case when only a single vertex is perturbed. The details of the $m=2$ case are given in the respective section.
For laying out the resulting grid, we choose the following (arbitrary) convention to place the vertices: In the mapped graph the starting vertex $x_0$ is placed at the bottom left. The opposite corner of the HC, i.e., the bitwise complement $\overline{x_0}$, will also be the opposite corner in the grid and placed at the top right. The remaining otherwise distinguished vertex $x_q$ is placed at the bottom right in the grid. By symmetry, its bitwise complement $\overline{x_q}$ is placed at the top left of the grid. Note further that for symmetry considerations, it is only the Hamming weight $q$ of the perturbed vertex that plays a role and not the exact chosen vertex $x_q$.

The remaining vertices of the HC are grouped by symmetry and also placed in the grid. Their horizontal and vertical coordinates in the grid are defined by their bit string in the following way. The bit string of the perturbed vertex $x_q$ defines two sets of bits (symbols) in the string, namely, those bits that have bit value 1 in $x_q$ and those bits that have bit value 0. Let us call these sets $X$ for the 1-bits and $Y$ for 0-bits in $x_q$. There are therefore $\abs{X}=q$ bits in set $X$ and $\abs{Y}=d-q$ bits in set $Y$. To give an example, for $x_q=0011_2$ the set $X$ contains both lower bits with value 1 (bits 1 and 2 when counted from the right) and set $Y$ contains the two higher bits with value 0 (bits 3 and 4 when counted from the right).

The $x$ and $y$ coordinates of every vertex in the grid are given by the Hamming weight of its bits in subsets $X$ and $Y$ denoted by $\abs{\cdot}_X$ and $\abs{\cdot}_Y$, respectively. That is, $x_0$ has coordinate $(0,0)$ and its opposite corner has coordinate $(q,d-q)$. The distinguished vertex $x_q$ has coordinate $(q,0)$. Placed at coordinate $(1,0)$ are all $\binom{q}{1}=q$ vertices that have only a single bit with value 1 in their bit string, which is in one of the positions that belong to subset $X$. Similarly, placed at coordinate $(0,1)$ are all $\binom{d-q}{1}=d-q$ vertices with a single bit with value 1, which is in one of the positions of subset $Y$. The mapping maintains the relation that vertices that differ only in one bit are neighbors and connected in the grid and that from the starting vertex, $q$ steps need to be walked to reach the distinguished vertex $x_q$ and $d$ steps need to be walked to reach the opposite corner.

Generally, a 2D grid of $(q+1)\times(d-q+1)$ vertices emerges, where $\binom{q}{x}\binom{d-q}{y}$ vertices of the HC are mapped to position $(x,y)$ in the grid.
Figure~\ref{fig:mapping_illustration} illustrates the mapping.
Depending on the Hamming weight $q$ of the perturbed vertex $x_q$, the grid size ranges from  $1\times(d+1)$ vertices on a vertical or horizontal line for $x_q=x_0$ or $\overline{x_0}$ to the maximal grid size of $\tfrac{1}{4}(d+2)^2$ when $q=\tfrac{d}{2}$ for even $d$ and of size $\tfrac{1}{4}(d+1)(d+3)$ when $q=\tfrac{d\pm1}{2}$ for odd $d$.

On this 2D grid, the walker can only go to one of its four neighbors, that is, rightward ($R$) and leftward ($L$), which would correspond to increasing and decreasing its $x$ coordinate by one, respectively, and upward ($U$) and downward ($D$), which would amount to increasing and decreasing its $y$ coordinate, respectively. The probability 
$p_{xy}(J)=\tfrac{1}{N_{xy}(J)}$ to go from coordinate $(x,y)$ on the grid in direction $J\in \{R,L,U,D\}$ is then expressed by the inverse of the number of original edges that effectively lead to direction $J$, which at unperturbed vertices is given by
\be
\label{eq:specifying_N_for_tails}
N_{xy}(J) = 
\begin{cases}
	q-x, 	& J=R \\
    x,		& J=L \\
	d-q-y,	& J=U \\
	y,		& J=D \\
\end{cases}
\ee
up to small exceptions at the perturbed vertices, which reflect the change in degree and connectivity.  As a sanity check we note that summing over all directions gives the degree of the vertex ($d$ for the regular graph), as required.  
 
The resulting size of the reduced graph scales therefore only with $d^2$ in the worst case rather than with $2^d$ for the full HC. The mapping is applied to the graph on which the walk takes place and therefore simulations of both classical and quantum walks will benefit from this reduction in dimension.

\subsection{Mapping of the walk operator}

With the mapping of the HC to a 2D grid at hand, we can now identify the basis vectors for the quantum walk in the reduced effective subspace in which the walk takes place. 
The new basis states at the grid vertex $(x,y)$ with $x\in\{0,1,\dotsc,q\}$ and $y\in\{0,1,\dotsc,d-q\}$ in direction $J\in \{R,L,U,D\}$ can then be expressed in terms of the full basis states $\{\ket{j,x_i}\}$ as follows (see \cite{2014_HCI_PRA} for corresponding expressions in the case of local or symmetric embeddings):
\be
\label{eq:new_2D_basis}
\ket{J,x,y} =  \frac{1}{\sqrt{N(J,x,y)}} \sum_{\substack{x_i\in V,\\(\abs{x_i}_X=x,\\ \phantom{(}\abs{x_i}_Y=y)}} \sum_{\substack{j=1\\ (j\mapsto J)}}^{p_i}
	\ket{j,x_i},
\ee
where the outer sum runs over all vertices of the HC that are mapped to the same vertex with coordinates $(x,y)$ on the grid, the inner sum runs over all directions $j$ that are mapped to direction $J$ on the grid, and the normalization factor is given by
\be
\label{eq:normalization}
N(J,x,y)=\tilde{N}(x,y)N_{xy}(J), 
\ee
with $\tilde{N}(x,y) = {q\choose x}{d-q \choose y}$ the total number of HC vertices that are mapped to vertex $(x,y)$ on the grid and $N_{xy}(J)$ giving, for each such state, the number of original directions $\ket{j}$ that effectively lead to direction $J$, as given in \eqref{eq:specifying_N_for_tails}.
The shift operator, defined in Eq.~\eqref{eq:shift_operator}, is then given by 
\begin{align}
\label{eq:new_2D_shift_operator}
S =& \sum_{x=0}^{q-1} \sum_{y=0}^{d-q} \Big[\ket{L,x+1,y}\!\bra{R,x,y} + \mathrm{H.c.} \Big] \\ \nonumber
  &+ \sum_{x=0}^{q} \sum_{y=0}^{d-q-1} \Big[ \ket{D,x,y+1}\!\bra{U,x,y} + \mathrm{H.c.} \Big]
\end{align}
and the Grover coin operator defined in Eq.~\eqref{eq:coin_operator} is given by
\begin{equation}
\label{eq:new_2D_coin_operator}
C_d = \sum_{x=0}^{q} \sum_{y=0}^{d-q} \sum_{J,K} c(J,K,x,y) \ket{J,x,y}\!\bra{K,x,y},
\end{equation}
where the sum over $J$ and $K$ runs over all directions $\{R,L,U,D\}$ on the grid and the matrix elements are 
\begin{align}
\label{eq:coef_definition_for_coin}
c(J,K,x,y) &= \bra{J,x,y}C\ket{K,x,y} \nn \\
           &= 
\left\{ 
\begin{matrix}
\frac{2}{d}\sqrt{N_{xy}(J)N_{xy}(K)}, & \mbox{$J\!\neq\!K$} \\
\frac{2}{d}N_{xy}(J)-1,               & \mbox{$J\!=\!K$}.
\end{matrix} 
\right.
\end{align}

\begin{figure*}[p]	
  \includegraphics{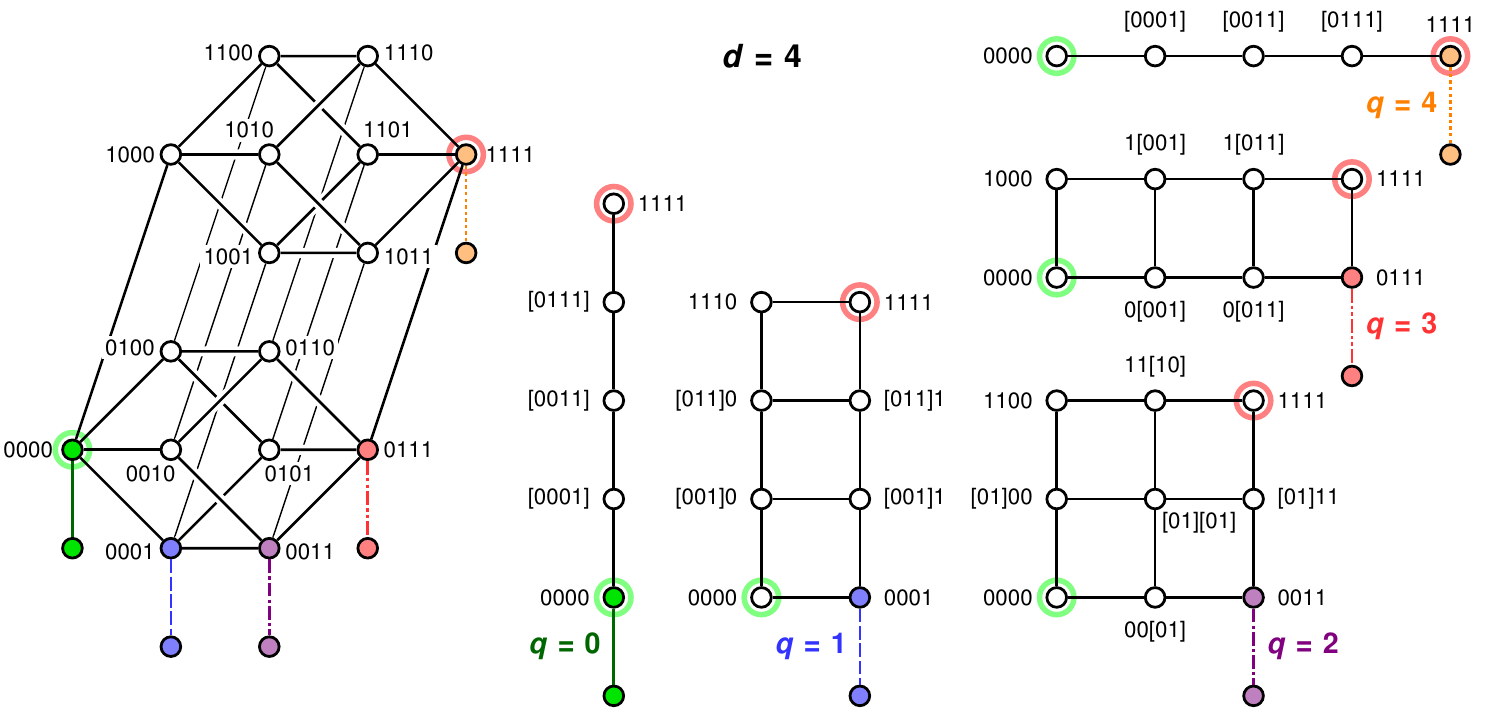}
  \caption{Added vertex: All nonequivalent cases for attaching a tail to a four-dimensional HC are shown with different colors and line types (left). Also shown is a complete family of respective mapped grid graphs (right). Start and end vertices of the random walk are circled in green and red; on the HC they are the leftmost and rightmost vertices and on the grid the bottom left and top right vertices, respectively. Labels of vertices on the grid represent those bit strings of the HC vertices that can be obtained by all permutations of symbols inside square brackets.}
  \label{fig:single_tail_illustration4D}
\end{figure*}

\begin{figure*}[p]
	\begin{minipage}[c]{0.53\linewidth}
		\includegraphics[width=\textwidth]{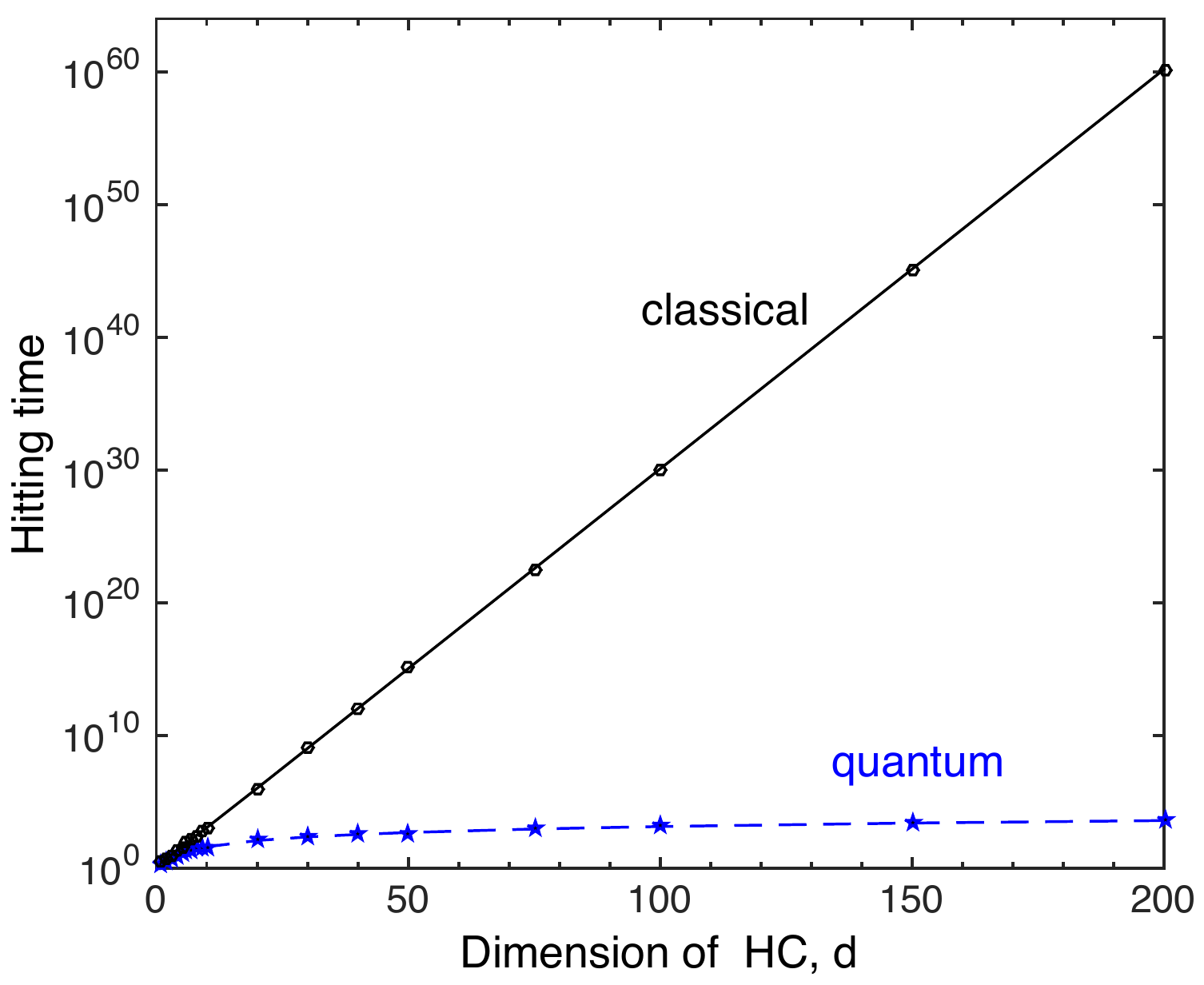}%
	\end{minipage}%
	\begin{minipage}[c]{0.465\linewidth}
	\includegraphics[width=\textwidth]{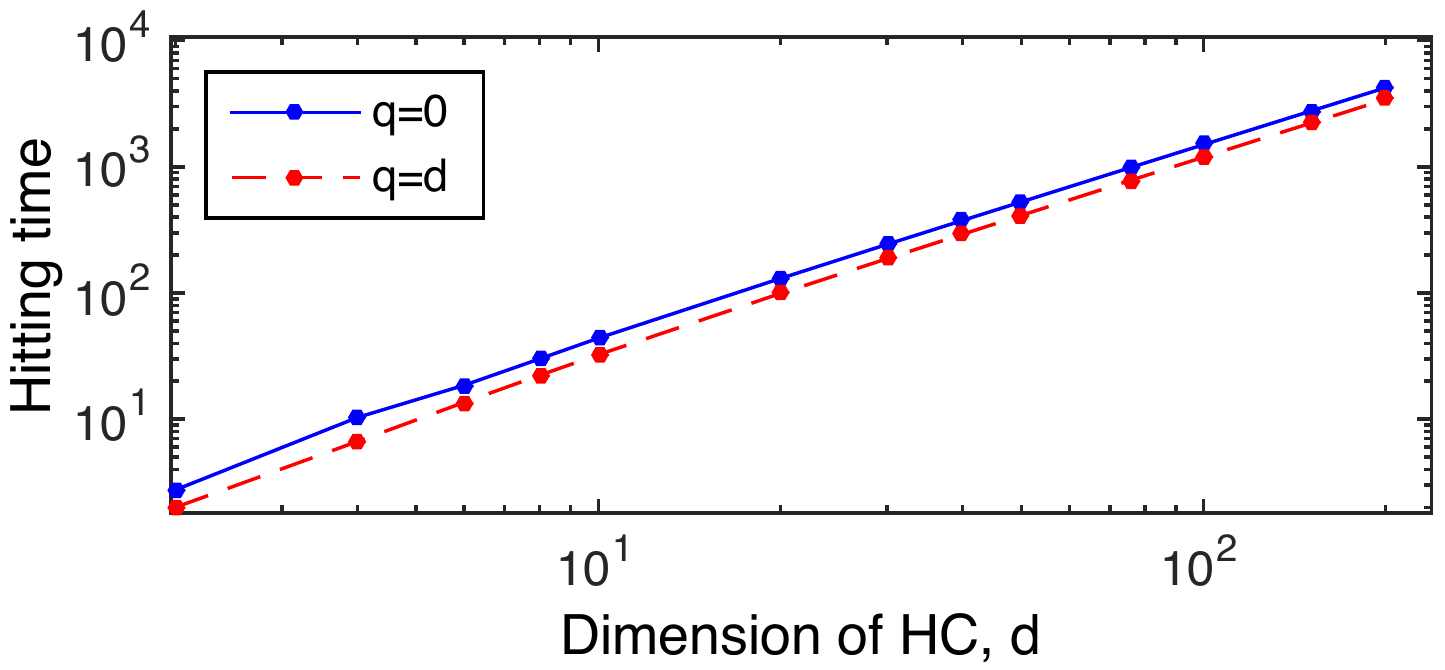}\\		
	\includegraphics[width=\textwidth]{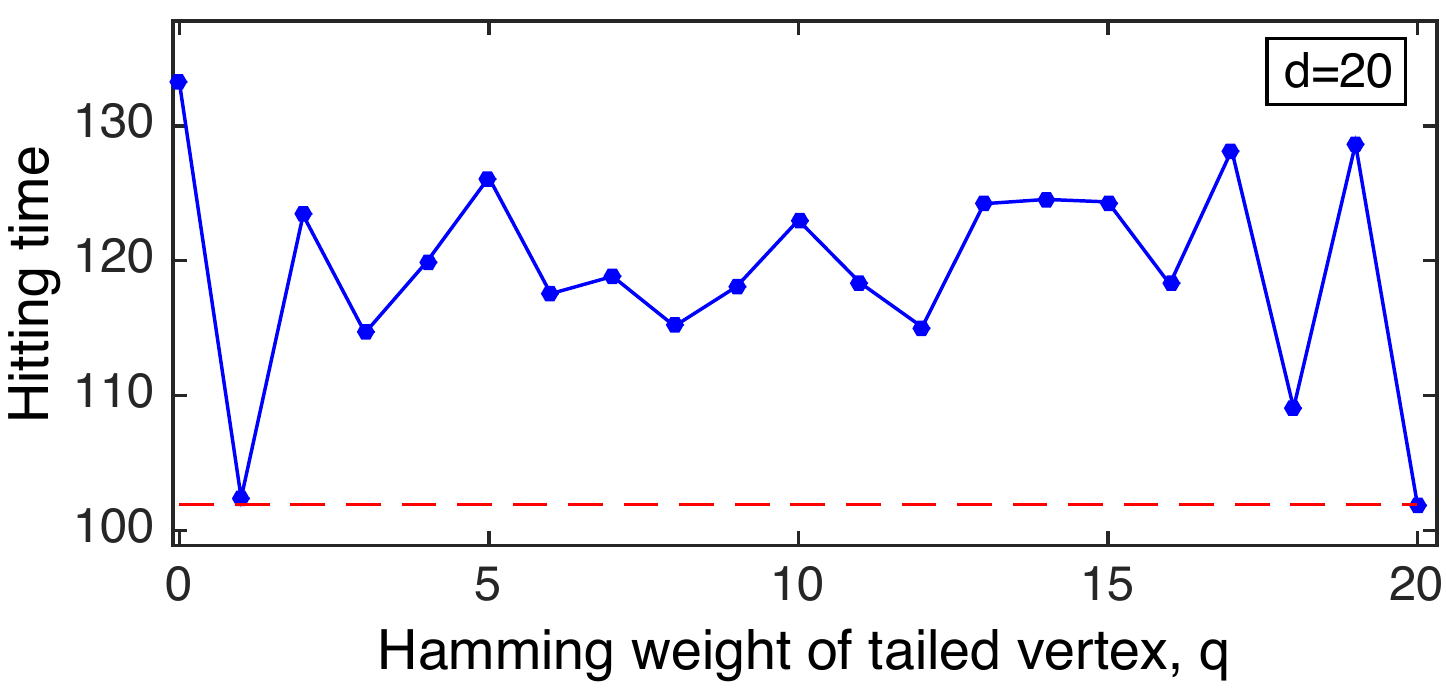}
	\end{minipage}%
	\caption{Added vertex: Expected hitting time of a walker on a $d$-dimensional HC with a single tail of length one connected to a vertex of Hamming weight $0 \leq q \leq d$.
		Shown on the left are classical and quantum walks with $q=d-1$. The classical data are calculated according to \eqref{eq:single_tail} and the quantum data are from numerical simulations with an error threshold of $\epsilon=10^{-4}$.
		Shown on the top right is the quantum walk for extreme cases of the expected hitting time with $q=0$ and $q=d$ (the latter being equivalent to the bare HC) and an error threshold of $\epsilon=10^{-4}$. The numerical fit yields the power $n\simeq 1.5$ for each curve.
		On the bottom right is a quantum walk with fixed $d=20$ and varying $0\leq q \leq 20$. The horizontal line indicates the quantum hitting time for $q=20$ (bare HC). The data are from simulations with an error threshold of $\epsilon=10^{-10}$.}
	\label{fig:tail_varying_d}
\end{figure*}

\section{Results} \label{sec:results}
\subsection{Nonsymmetric embedding (added vertex)} \label{secsec:non_sym_emb}

We first consider a nonsymmetric case of embedding the HC into a larger graph. A minimal modification of this kind is the case of a single tail of length one, i.e., a single additional vertex, connected to one of the vertices of the HC. For the mapping, the perturbed vertex in the HC is the vertex to which the additional graph (the additional vertex) is appended. Due to symmetry only the Hamming weight $q$ of the tailed vertex plays a role. For illustration of the mapping procedure we provide in Fig.~\ref{fig:single_tail_illustration4D} the mapping of a single tail attached to a HC of dimension $d=4$ for all possible Hamming weights, that is, with $0 \leq q \leq d$. Note that when $q=d$ we revert back to a walk on the bare HC as the walker never reaches the appended vertex because it is only accessible through the final vertex. The mapping of the walk operator is exactly as described in Sec.~\ref{sec:mapping} except for the vertex $x_q$ and the added vertex. The degree of both has to be taken into account explicitly by associating a coin of dimension $p=1$ with the added vertex, and a coin of dimension $p=d+1$ with the vertex $x_q$, that is, by applying a conditional coin operator, according to Eq.~\eqref{eq:conditioned_coin_operator}. The corresponding changes of $N_{xy}(J)$ are $N_{q0}(D)=1$ for the modified vertex $x_q$, and for the added tail vertex the only nonzero value is $N_{\text{tail}}(U)=1$.

To obtain the expected hitting time for the quantum walk in the single-tail case, we performed numerical simulations of the walk on the effective reduced subspace. 
This reduction in dimension allows us to consider hypercubes of high dimensions (of the order of hundreds) as shown in Fig.~\ref{fig:tail_varying_d}. 

In the classical case, when the HC is locally appended (not necessarily symmetric) with additional graphs, i.e., the HC is embedded in a larger graph, the expected hitting time to reach the opposite corner on this HC was shown to be \cite{2014_HCI_PRA}
\be
\label{eq:embedded_hcI}
\tau(0) = \sum_{k=0}^{d-1}  \frac{\sum_{j=0}^{k} {d\choose k-j}\alpha_{k-j}} {{d-1\choose k}},
\ee 
where $\alpha_k = \frac{e_k}{d}+1$ and $e_k$ is the average number of edges starting at the vertices in all the external graphs that are attached to HC vertices of Hamming weight $k$ (including the edges connecting the HC and the attached graph, so-called legs).
Note that when $e_k=0$ for all $0 \leq k \leq d-1$, the expression of Eq.~\eqref{eq:embedded_hcI} correctly reduces to the expression of Eq.~\eqref{eq:classical_ht_hc}.
By virtue of Eq.~\eqref{eq:embedded_hcI}, the expected hitting time for the classical walk on a HC of dimension $d$ with a single additional vertex attached to a HC vertex of Hamming weight $q<d$ is given analytically by
\begin{equation}
\tau^{\text{tail}}_{q,d} = \sum_{k=0}^{d-1}\frac{ \sum_{j=0}^{k} {d\choose k-j} + \sum_{j=0}^k {d\choose k-j}\frac{e_{k-j}}{d}} {{d-1\choose k}},
\end{equation}
where $e_k=0$ for $k\ne q$ and $e_q = 2/{d\choose q}$ (the factor of 2 is due to taking both the ingoing and the outgoing directions of the additional edge into account).
We can separate the hitting time for the walk on the regular HC and obtain
\begin{equation}
\label{eq:single_tail}
\tau^{\text{tail}}_{q,d} = \tau(0) + \frac{2}{d}\sum_{k=q}^{d-1} \frac{1}{{d-1\choose k}} = \tau(0) + O\left(\frac{1}{d}\right),
\end{equation}
where the leading term in the correction is $4/d$ for $q=0$ and $2/d$ otherwise. The correction is exactly $2/d$ for $q=d-1$.
From \eqref{eq:single_tail} it is readily apparent that the expected hitting time of the classical walk is monotonically decreasing with the Hamming weight of the tailed vertex $q$ and that the perturbation in terms of the expected hitting time decreases with the HC dimension~$d$.

Figure~\ref{fig:tail_varying_d} (left) shows a comparison of the expected hitting times of the classical and quantum walks of dimension $d$, with a single tail attached to a vertex of Hamming weight $q=d-1$. The comparison on a log-plot illustrates that the gap between the classical and the quantum curves increases exponentially with the HC dimension $d$, implying that the exponential speed-up of the quantum walk compared to its classical counterpart survives also when the overall structure is not fully symmetric. 

We observe that for both the classical and quantum walks the precise values of the expected hitting time depend on the Hamming weight $0\leq q \leq d-1$ of the tailed vertex, but they are of the same order of magnitude for all~$q$.
Figure~\ref{fig:tail_varying_d} (top right) illustrates that the expected hitting time of the quantum walk appears to scale with $d$ according to a power law, i.e., $d^n$ with $n$ constant. Both curves display extremal cases of the hitting time, the longest for the tail attached directly to the starting vertex $q=0$ and the shortest for the tail attached to the final vertex, which is equivalent to a walk on the bare HC. A numerical fit~\cite{EZYFIT_MATLAB} yields a power of $n\simeq 1.5$ for each curve.
Figure~\ref{fig:tail_varying_d} (bottom right) displays for one example of $d=20$ that, numerically, we obtain a nontrivial dependence of the exact value of expected hitting time on $q$, which is neither monotonic nor particularly symmetric.
We can rule out convergence issues in our simulation to the extent that we used an error threshold of $\epsilon=10^{-10}$.
Qualitatively, however, the apparent scaling and order of magnitude of the hitting time is largely independent of $q$, implying that an exponentially growing gap between expected hitting times of quantum and classical walks exists for all $q$.

\subsection{Nonlocal embedding (HC inside HC)} \label{secsec:non_local}

\begin{figure}[tb]
  \includegraphics{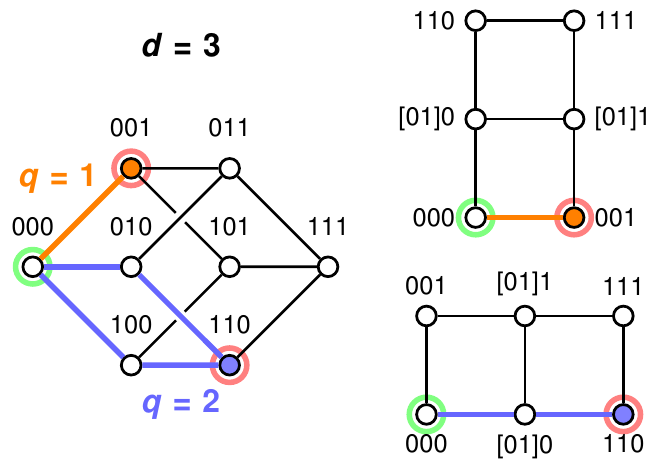}
  \caption{HC inside HC: Nonlocal embedding of two low-dimensional examples of HCs (dimension $q=1,2$) as subgraphs in a three-dimensional HC (left) and the corresponding mapped graphs (right). Edges and final vertices of the lower-dimensional HCs are marked in the same color.
  }
  \label{fig:hc_in_hc_illustration}
\end{figure}

Next we consider a nonlocal embedding of a HC in a larger graph.
An extreme case of a nonlocal embedding is obtained when every vertex of the embedded HC is attached to an external graph in such a way that any two vertices of the embedded HC are connected through the external graph. Such cases are obtained when a HC of dimension $q$ is embedded as a subgraph in a larger HC of dimension $d>q$, as demonstrated on the left side of Fig.~\ref{fig:hc_in_hc_illustration}. Effectively, this embedding is obtained when the target vertex of the walk is not the opposite corner on the $d$-dimensional HC, but rather a different vertex of Hamming weight $1 \leq q < d$.

When investigating the scaling behavior of the expected hitting time, we need to keep in mind what exactly it is that should scale. In this section we are interested in the walk on the hypercube of dimension $q$, i.e., we primarily investigate the scaling with respect to $q$. The dimension $d$ of the larger HC in which our primary HC is embedded may either be just larger by a constant, e.g., $d=q+1$, or scale with $q$, e.g., by a factor such as $d=2q$. Note that both variants of embedding the HC in a larger HC differ in the way that the embedding is constructed. For a constant difference (e.g., $d=q+1$) the larger HC is obtained by a fixed number of duplications of the primary HC,\footnote{A HC of dimension $d+1$ can be recursively constructed, by duplicating a HC of dimension $d$ and connecting each vertex from the original HC with its cloned one.} whereas for a difference that monotonically increases with $q$ the embedding procedure itself grows, which leads to qualitatively different scaling behaviors of the hitting time.

The mapping of the HC results in a 2D grid as detailed in Sec.~\ref{sec:mapping}. The distinguished vertex in our mapping is given by the final vertex of the lower $q$-dimensional HC (see Fig.~\ref{fig:hc_in_hc_illustration}). The vertex $\overline{x_0}$ opposite the starting vertex is only distinguished due to symmetry but is otherwise identical to the other vertices. The new final vertex $x_q$ is the place where the walker is absorbed and the projective measurement is applied in the quantum walk.

For the classical walk, an analytical expression for the expected hitting time can be derived with the following argument. First, note that the expected hitting time to reach a vertex of Hamming weight $q$ when starting from $x_0$ equals the expected hitting time to reach a vertex of Hamming weight $d$ when starting from a vertex with Hamming weight $d-q$. In both cases the starting vertex and final vertex are $q$ steps apart on the HC. Symmetry dictates that the expected hitting time depends only on the relative Hamming distance, i.e., $\tau_{x_0 x_q} = \tau_{x_{d-q} x_d}\equiv\tau(d-q)$. Then, by following the same procedure as given in \cite{2006a_Krovi_PRA} we obtain
\begin{align}
\label{eq:hc_in_hc_classical1}
\tau_{q,d}^{\text{HC in HC}}
	&= \tau(d-q)=\sum_{k=d-q}^{d-1}\Delta(k) \\
\label{eq:hc_in_hc_classical}
	&=\sum_{k=d-q}^{d-1}\frac{\sum_{j=0}^{k} {d\choose k-j}} {{d-1\choose k}},
\end{align}
where $\Delta(k)$ is defined after Eq.~\eqref{eq:classical_ht_hc}. Note that when $q=d$ this expression reduces to the classical hitting time of the bare $d$-dimensional hypercube as given in~\eqref{eq:classical_ht_hc}.
In the sum in \eqref{eq:hc_in_hc_classical1} $d-q$ is the starting index of the summation over positive quantities. Therefore, the expected hitting time of the classical walk increases monotonically with~$q$. Moreover, using the recursive formula for $\Delta(k)$~\cite{2006a_Krovi_PRA},
\be
\label{eq:recursive_on_Delta}
\Delta(k) = \frac{d\!-\!k\!-\!1}{k\!+\!1}\Delta (k\!+\!1)-\frac{d}{k\!+\!1}, 
\ee
it can be verified that for all $k<d-1$, $\Delta(k) < \Delta(d-1)$,
which implies that $\tau_{q,d}^{\text{HC in HC}}$ is upper bounded by $q \tau_{q=1,d}^{\text{HC in HC}}$.

For the quantum walk we first need to reconsider the definition of the expected hitting time. We observe that whenever $q\neq 1,d-1,d$ the walker remains trapped inside the graph and cannot reach the final vertex with finite probability (a phenomenon that was already indicated in~\cite{2014_HCI_PRA}). This inability to reach the final vertex is clearly due to the changed symmetry of the walk and the same was observed in differently distorted HCs or when using a different coin~\cite{2006a_Krovi_PRA,2006b_Krovi_PRA,2008_Krovi_PRA}. Under the dynamics driven by the walk operator, part of the walker always interferes destructively on the final vertex and thus remains on the graph forever.
The total probability to reach the final vertex at all is
\begin{equation}
	p_\text{tot} = \sum_{t=0}^\infty p_{x_f}(t) < 1,
\end{equation}
i.e., it no longer sums to one.
Note that a trapped walker never occurs in the classical walk on connected graphs. For a meaningful comparison we therefore consider the conditional expected hitting time for the quantum walk, that is, the expected hitting time given that the walker arrives at all at the final site. Analogous to the expected hitting time, we define $\tau_c = \sum_{t=0}^\infty t\, p_{x_f}(t \vert x_f)$, where $p_{x_f}(t \vert x_f)$ denotes the probability to hit the final vertex for the first time at time $t$ given that it hits the final vertex at all during the walk. By definition of conditional probabilities $p(A\vert B)=p(A \cap B)/p(B)$ and identifying the joint event ``to hit for the first time at time $t$ and to hit at all'' with ``to hit for the first time at time $t$'' we arrive at the expression for the conditional expected hitting time
\begin{equation} \label{eq:conditional_hitting_time}
	\tau_c = \frac{1}{p_\text{tot}} \sum_{t=0}^\infty t\, p_{x_f}(t)
\end{equation}
and $\tau_c$ reduces to the ordinary expected hitting time for $p_\text{tot}=1$.
Since the quantum walk reaches the final vertex only with probability $p_\text{tot}$, one needs to either restart the walk $1/p_\text{tot}$ number of times or run this number of walks in parallel to observe (on average) that one walker hits the final vertex.

The trapping of the walker inside the HC can be captured more formally. See~\cite{2006b_Krovi_PRA} for a detailed analysis in terms of symmetries of the graph, of which we repeat the following properties that allow, in principle, for a computation of $p_\text{tot}$ from the eigenvectors of $U$ for a given $x_f$. The walk starts in state $\ket{\psi_0}$ and evolves under multiple applications of the walk operator $U\Pi_0$. The fact that it is trapped inside the HC and that equivalently, in analogy to a similar phenomenon in quantum optics~\cite{Fleischhauer}, it evolves into a dark state implies that the state has support in a subspace of dark states. This subspace $\mathcal{V} \subset \mathcal{H}$ has the following properties: (i) $\mathcal{V}$ has no overlap with the final state $\braket{\psi_f}{v}=0\, \forall \ket{v}\in\mathcal{V}$ and therefore the projection $\Pi_0$ is the identity operation when restricted to this subspace and (ii) $\mathcal{V}$ remains invariant under applications of the walk operator $U\Pi_0\ket{v}\in\mathcal{V}\, \forall \ket{v}\in\mathcal{V}$. Both properties together imply that $\mathcal{V}$ is also invariant under applications of $U$ alone, that is, $U$ is block diagonal with respect to dark states and nondark states (which form the orthogonal complement of the space of dark states). From this invariance we obtain that the overlap of $\ket{\psi_t}=(U\Pi_0)^t\ket{\psi_0}$ with the dark space $\mathcal{V}$ is time independent:
\begin{align}
	\label{eq:overlap_dark_space_is_time_independent}
	\bra{\psi_t}\Pi_V\ket{\psi_t}
	&= \sum_v \abs{\bra{v}(U\Pi_0)^t\ket{\psi_0}}^2 \nonumber \\
	&= \sum_v \abs{\braket{v}{\psi_0}}^2,
\end{align}
where $\Pi_V=\sum_v \proj{v}$ with $\{\ket{v}\}_v$ being a basis in $\mathcal{V}$ 
and the walk operator $U\Pi_0$ is applied to the left.
Since the overlap with the space of dark states does not change with time, the part of the walk that is locked in the graph is given by the overlap of the initial state with all dark states:
\be
	\label{eq:overlap_with_dark_space}
	1-p_\text{tot} = \sum_v \abs{\braket{v}{\psi_0}}^2.
\ee
The probability $p_\text{tot}$ can therefore be calculated using the eigenvectors of~$U$ since a basis of $\mathcal{V}$ can be chosen from eigenvectors of~$U$.

\begin{figure}[tb]
	\centering
	\includegraphics[width=\linewidth]{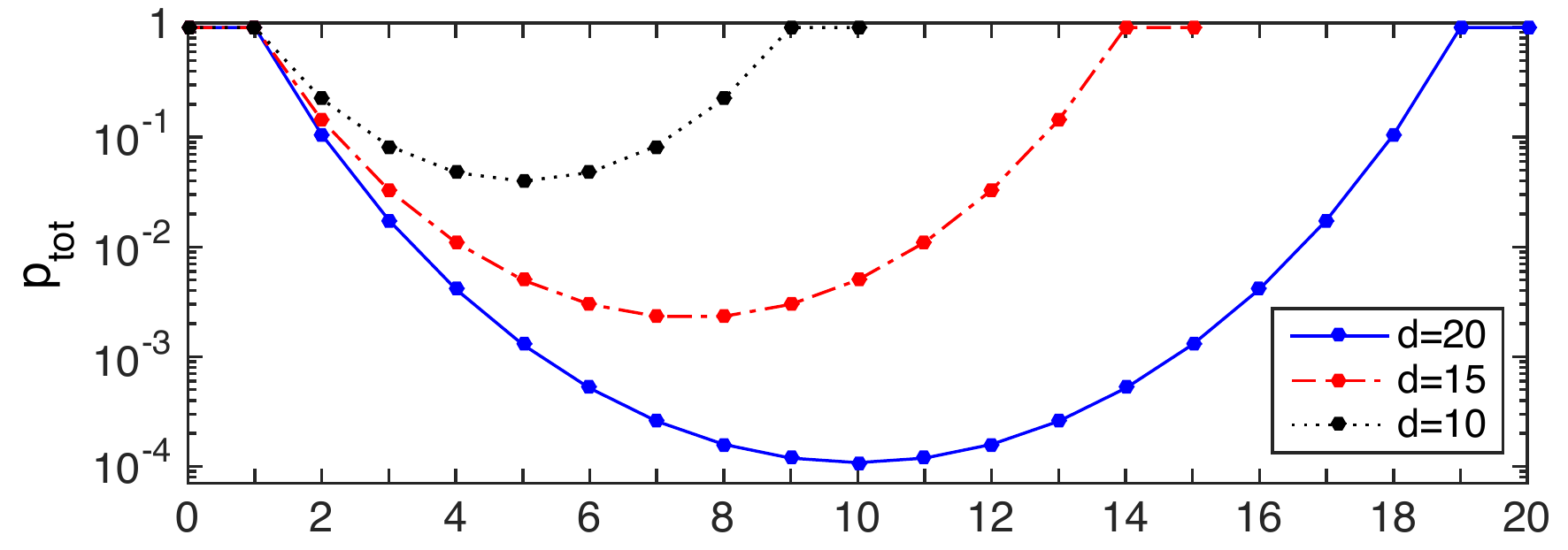}
	\includegraphics[width=\linewidth]{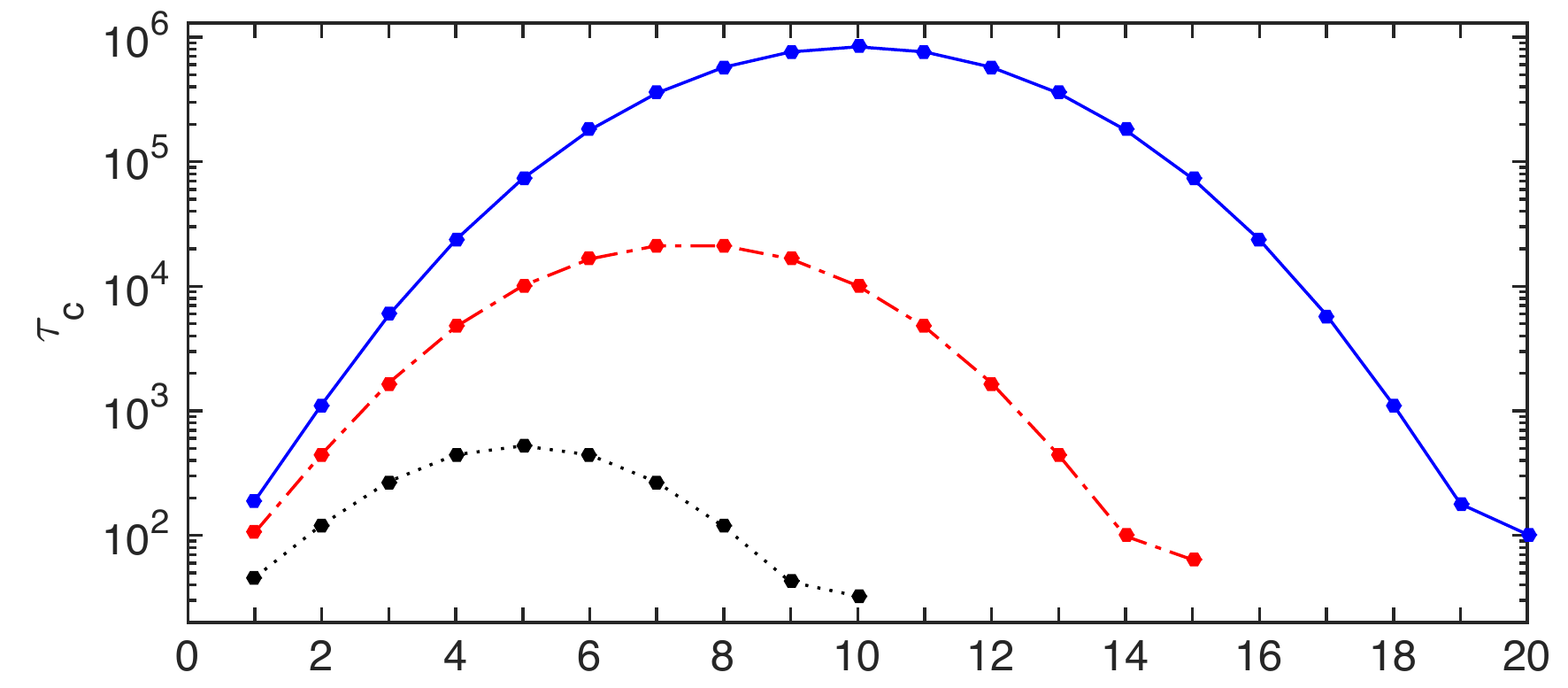}	
	\includegraphics[width=\linewidth]{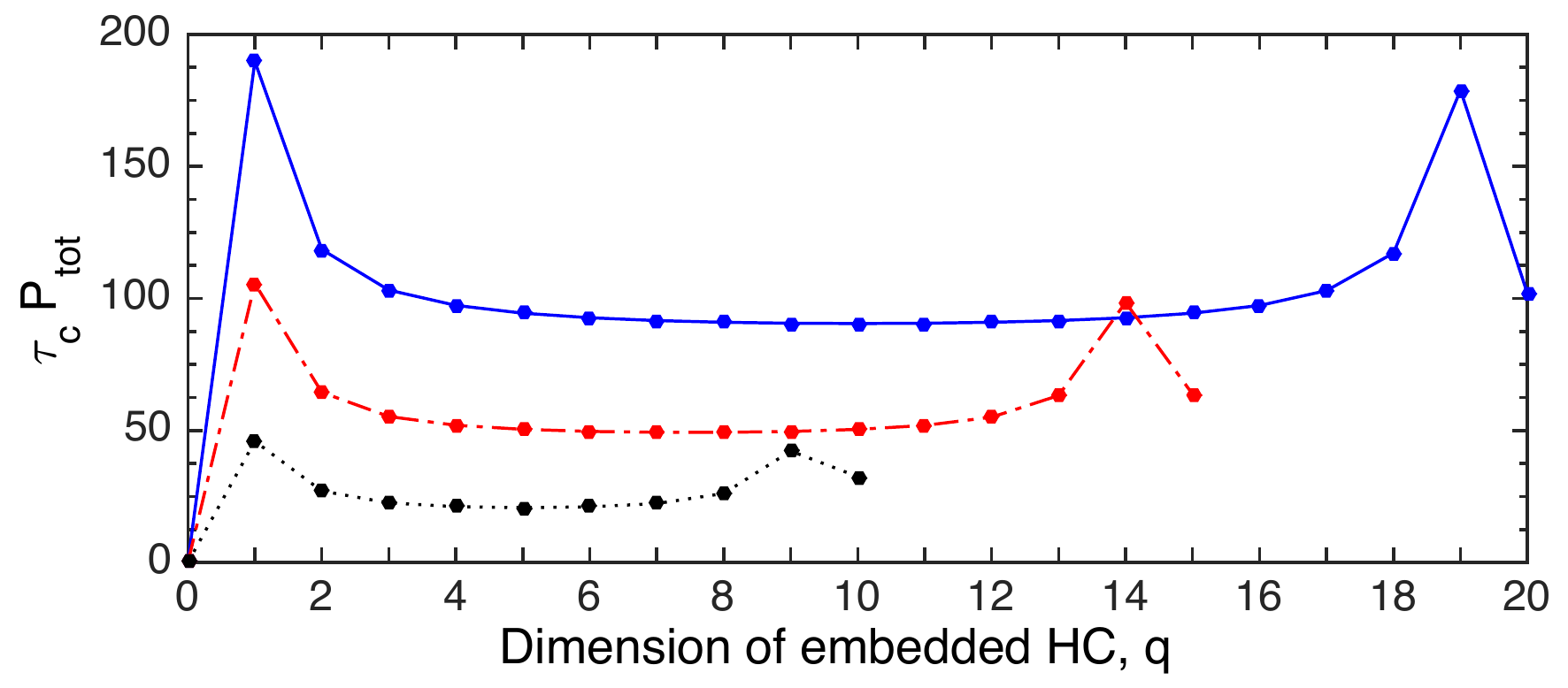}
	\caption{HC inside HC: Total probability to reach the final vertex at all during a single walk $p_\text{tot}$ (top), conditional hitting time $\tau_c$ (middle), and their product (bottom), i.e., the numerator in \eqref{eq:conditional_hitting_time}, depending on $q$ (Hamming weight of final vertex) for $d=10,15,20$. The data are from numerical simulations with corresponding error thresholds $\delta=10^{-5}, 10^{-6}, 10^{-7}$ and a time window parameter $t_W=\delta^{-1}$. For points without dark states ($q=1,d-1,d$) an error threshold of $\epsilon=10^{-5}, 10^{-6}, 10^{-7}$ was employed for $d=10,15,20$, respectively.}
	\label{fig:hc_in_hc_ptot_and_tau_c}
\end{figure}

Figure~\ref{fig:hc_in_hc_ptot_and_tau_c} (top) shows that $p_\text{tot}=1$ only for $q=1,d-1,d$ and that $p_\text{tot}$ is symmetric around $q=d/2$, where it attains a minimum. The logarithmic plot illustrates that $p_\text{tot}$ is exponentially suppressed for values of $2\le q\le d-2$, and the parabolic shape suggests an exponential suppression in $q$ with exponent $O\big(-(q-\frac{d}{2})^2\big)$.
The conditional hitting times in Fig.~\ref{fig:hc_in_hc_ptot_and_tau_c} (middle) are of similar magnitude for values of $q=1,d-1,d$ and they are exponentially longer for $2\le q \le q-2$ with a maximum at $q=d/2$. The parabolic shape of the conditional expected hitting time on the logarithmic plot suggests an exponential increase in $q$ with an exponent $O\big((q-\frac{d}{2})^2\big)$.
Comparing the conditional hitting times for various $d$, it is apparent that $\tau_c(q)$ at $q=d-1$ scales differently with $d$ than the maxima of $\tau_c(q)$ at $q=d/2$. This difference in the scaling with $d$ is partly due to the scaling of the embedding procedure itself (the dimension $d$ of the full HC) when increasing the dimension $q$ of the principal HC that is embedded in the full HC.
Comparing the top and middle graphs of Fig.~\ref{fig:hc_in_hc_ptot_and_tau_c}, one realizes that the primary source of the exponential increase of $\tau_c$ is the exponential increase of factor $1/p_\text{tot}$ as illustrated in the top graph.

The fact that $p_\text{tot}<1$ has additional consequences for our numerical simulations, as calculating the conditional hitting time $\tau_c$ becomes delicate. This is because the infinite sums in Eq.~\eqref{eq:conditional_hitting_time} must still be truncated, to be numerically evaluated, but setting a small enough $\epsilon$ threshold such that $p_\text{tot}=(1-\epsilon)\rightarrow 1$, as explained in Sect.~\ref{secsec:quantum_prliminaries}, is no longer meaningful and merely results in an infinite calculation. Instead, the truncation should be done once it is safe to assume that the walk is already locked in the HC.
In practice, we verify that the probability to hit the target state does not increase by more than  $\delta \rightarrow 0$ during a long time window $t_Wd$, which scales with the HC dimension. To that end we decrease $\delta$ and increase the time window parameter $t_W$ until the conditional hitting time $\tau_c$ is well converged.

\begin{figure}[tb]
	\centering
	\includegraphics[width=\linewidth]{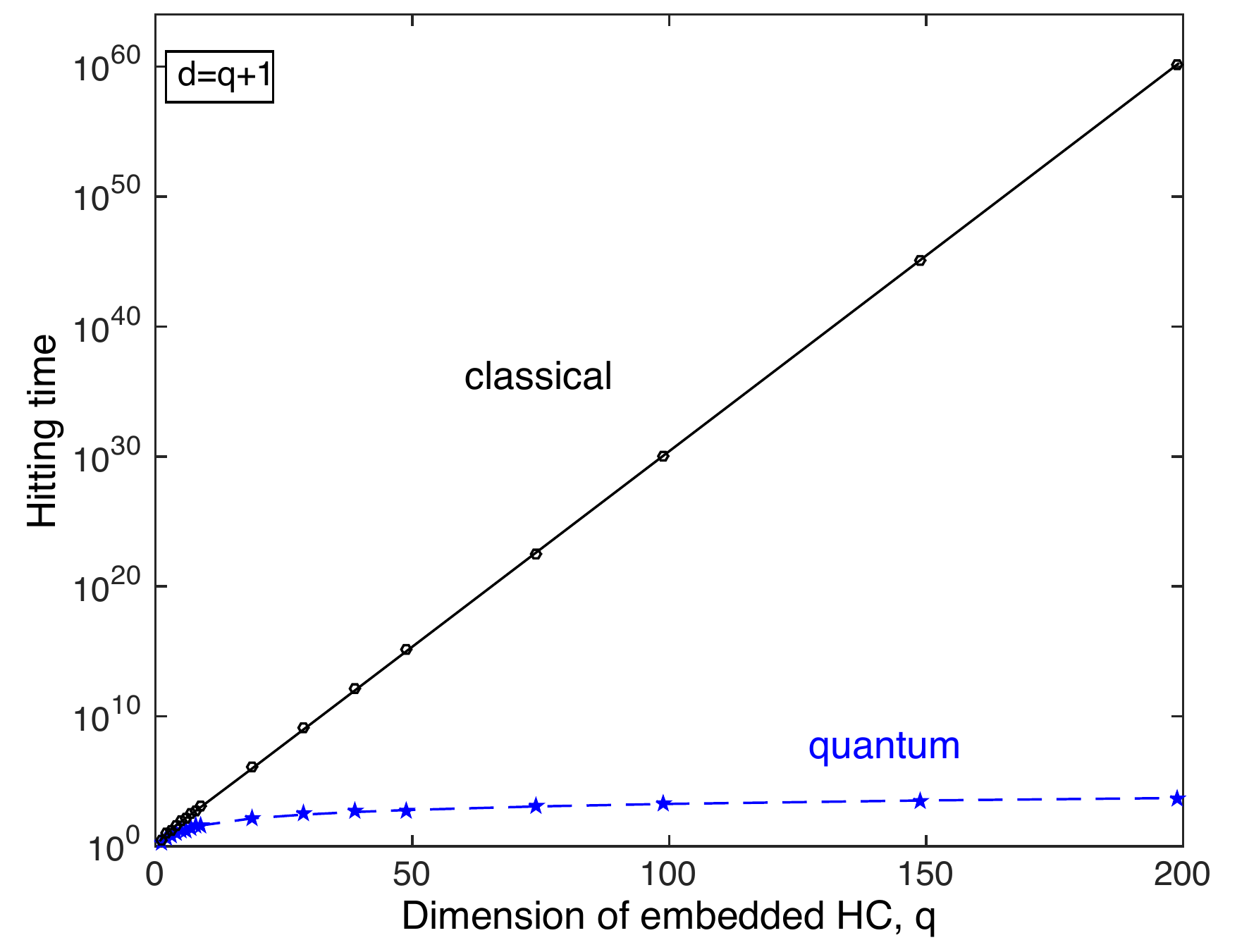}
	\includegraphics[width=\linewidth]{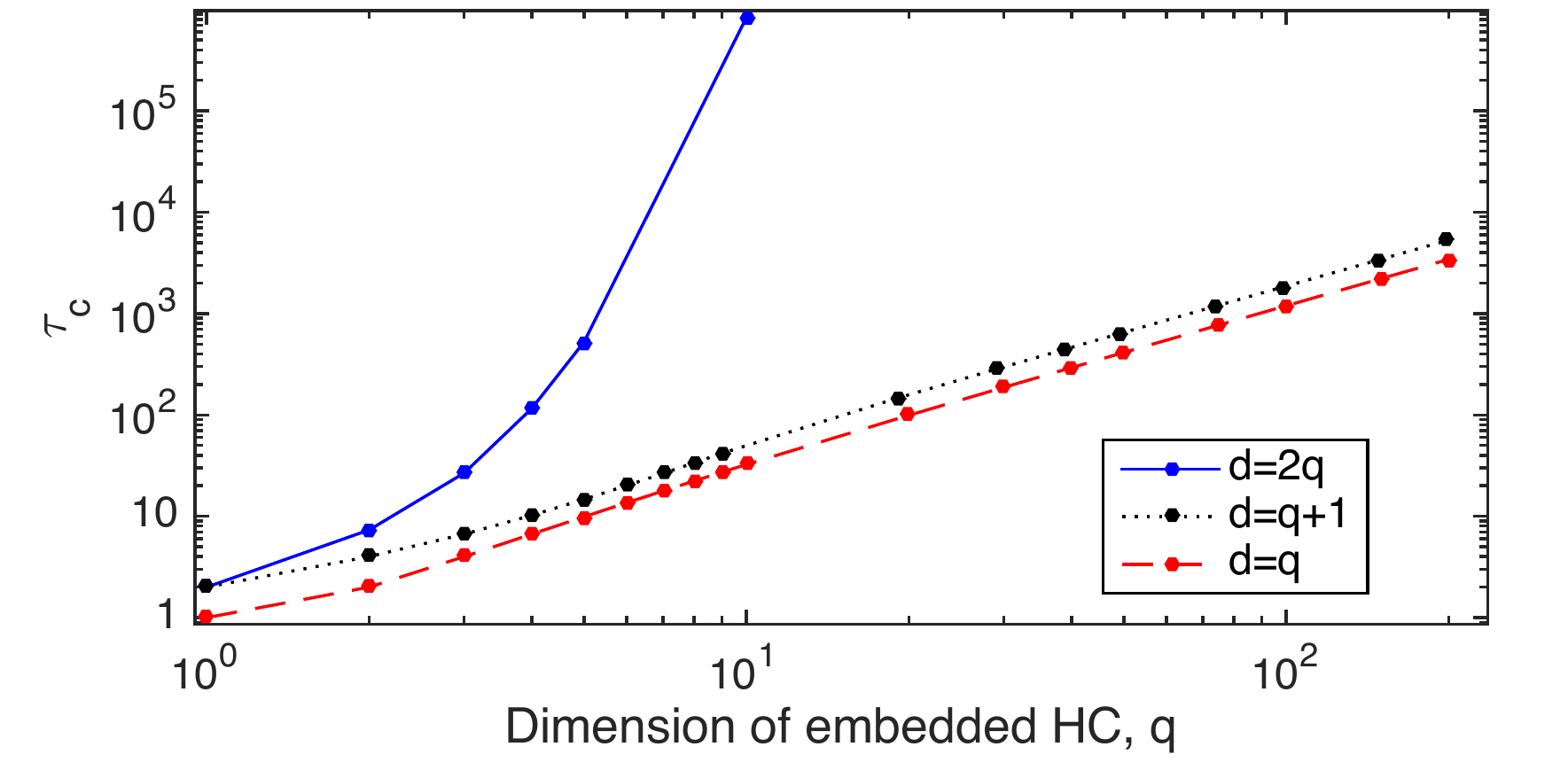}
	\caption{HC inside HC: Expected hitting times of a walker to hit the opposite corner of a $q$-dimensional HC that is embedded in a $d$-dimensional HC.
	Shown on top is a comparison between the classical and the quantum walks for the case $d=q+1$ (constant embedding). The quantum data are from numerical simulations with an error threshold $\epsilon=10^{-4}$.
	The bottom shows a comparison of conditional hitting times of quantum walks with $d=2q$ (embedding grows with $q$, solid blue line), $d=q+1$ (constant embedding, dotted black line), and $q=d$ (the bare HC, dashed red line) on a log-log scale. The data are from simulations with an error threshold of $\delta=10^{-7}$ and time window parameter $t_W=\delta^{-1}$ for the case $d=2q$ and an error threshold of $\epsilon=10^{-4}$ for the cases $d=q+1$ and $d=q$. The numerical fit to each of the lower curves ($d=q+1$ and $d=q$) yields $n\simeq 1.5$, whereas a numerical fit to the upper curve ($d=2q$) yields an exponential scaling with coefficient $\simeq\!1.4$.}
	\label{fig:hc_in_hc_varying_d}
\end{figure}

Figure~\ref{fig:hc_in_hc_varying_d} (top) shows a comparison of the expected hitting times for classical and quantum walks on a HC within a HC. As a benchmark we choose the case $d=q+1$, where the expected hitting times are plotted as a function of $q$. The figure illustrates that the difference between the classical and the quantum curves grows exponentially with $q$ and $d$, therefore implying that the exponential speed-up of the quantum walk on the HC remains also when the HC is embedded in a HC whose dimension is higher by one. 
The speed-up when increasing $d$ survives also when $q=1$ (not shown).

It is further observed that the expected hitting times of the quantum walk for cases without dark states ($q=1,d-1,d$) are in reverse order of the distance between the origin and the target vertices (the closer the slower), i.e., $\tau(d) < \tau(d-1) < \tau(1)$.
This is in contrast to the classical case, where the expected hitting time grows with the distance to the target vertex (the closer the faster) and in particular $\tau(d) > \tau(d-1) > \tau(1)$.
Figure~\ref{fig:hc_in_hc_varying_d} (bottom) illustrates that the quantum walk with a target vertex next to the opposite corner of the starting vertex maintains a longer hitting time when increasing $q$ and $d$, compared to the walk on the bare HC ($d=q$). Both curves appear approximately linear in the log-log plot, suggesting a power-law scaling with $q^n$, and a numerical fit yields $n\simeq 1.5$ for each curve. The case $d=2q$, in which the embedding procedure also grows with $q$, is shown in the same plot for comparison and it exhibits a qualitatively different behavior. The curve with $d=2q$ fits better to an exponential model and a numerical fit yields a scaling with $\sim\!\exp(1.4 q)$. These numerical results suggest that the quantum walk with $d=2q$ also falls into a class of walks with exponentially growing expected hitting times, like the classical walk.

In summary, we observe that an exponentially growing gap between the expected hitting times of quantum and classical walks persists for cases with $q=1,d-1$ and gives the same order of magnitude as the walks on the bare HC. Quantum walks with a different $q$, in particular with $q=d/2$, exhibit dark states and the conditional hitting time shows a qualitatively different scaling, which appears to be exponential as for the classical walk.

\subsection{Removing a single edge from the HC} \label{secsec:removed_edge}
\begin{figure}[tb]
  \includegraphics{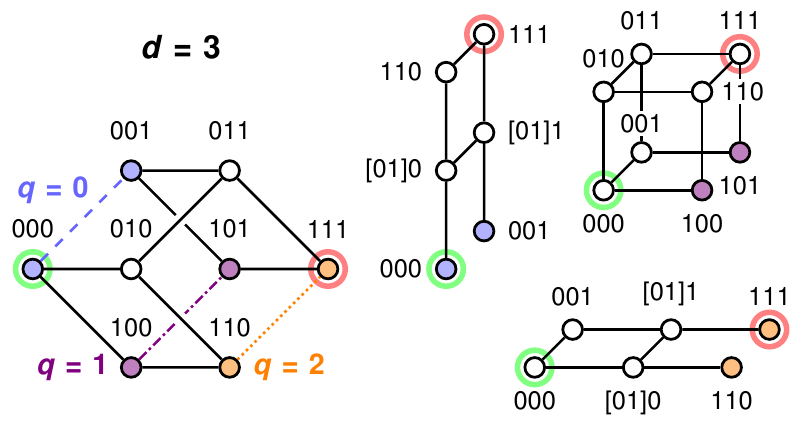}
  \caption{Removed edge: All equivalent cases of removing a single edge from a three-dimensional hypercube (left) together with their respective mapped graphs (right). Removed edges and thereby distinguished vertices are shaded with the same color. Starting and final vertices are circled as before.
  }
  \label{fig:edge_removal_illustration}
\end{figure}

In what follows we study the influence of removing a single edge from the HC on the resulting corner-to-corner hitting time. The resulting deficient HC can still be mapped to a grid of lower dimension, as depicted in Fig.~\ref{fig:edge_removal_illustration} and explained next.

Contrary to the cases we saw previously (of a HC with a single tail and a HC embedded inside a larger HC), removing a single edge from the HC distinguishes not just a single vertex of the HC, but two vertices, namely, those two neighboring vertices between which the edge is removed. We denote these distinguished vertices by $x_q$ and $x_{q+1}$. The additionally perturbed vertex, say, $x_{q+1}$, adds another dimension to the resulting mapped grids, which are therefore in general three dimensional.

The layout and coordinates of the resulting grid are obtained as follows. By construction, the bit strings of the two distinguished vertices differ only in the value of a single bit. This bit forms a separate set $Z$. The value of this bit gives the coordinate on the grid in the newly added (third) dimension. The remaining $d-1$ bits are equal for both vertices and define the subsets $X$ and $Y$ as before with all bits with value 1 forming a set $X$ of size $\abs{X}=q$ and all bits with value 0 forming a set $Y$ of size $\abs{Y} = d-q-1$. The resulting, in general three-dimensional, grid has two layers in the $z$ direction and a horizontal and a vertical extension of $q+1$ and $d-q$ vertices, respectively.
Accordingly, the number of vertices in the mapped grid graph is $2(q+1)(d-q)$ and ranges from its minimum of $2d$ vertices for an edge removed adjacent to the starting vertex ($q=0$) or the final vertex ($q=d-1$) to its maximum of $d(d+2)/2$ vertices for even $d$ and $(d+1)^2/2$ for odd $d$, respectively, when an edge is removed from the middle of the HC [at $q=d/2$ or $d/2-1$ for even $d$ and $q=(d-1)/2$ for odd $d$].
For a three-dimensional HC, a reduction in dimension is only obtained for $q=0,2$, but for larger $d$ a reduction in dimension is always achieved.
In the worst case we retain a structure of size that scales with $d^2$ instead of $2^d$ for the full HC.

When mapping the walk operator to the grid there are two new directions, forwards ($F$) and backwards ($B$), to move along the $z$ direction. The $N_{xyz}(J)$ are straightforwardly extended analogous to the $x$ and $y$ directions with a dependence on the $z$ coordinate. Only for the two perturbed vertices $x_q$ and $x_{q+1}$ a movement in the $z$ direction is not possible.

In the present case of a removed edge we lack a closed analytical expression for the corresponding expected hitting time of the classical walk. Therefore, we compute exact expected hitting times by solving the set of linear equations for the relevant entries of the fundamental matrix (see Sec.~\ref{secsec:classical_prliminaries}).

\begin{figure*}[t]
	\begin{minipage}[c]{0.55\linewidth}
		\includegraphics[width=\textwidth]{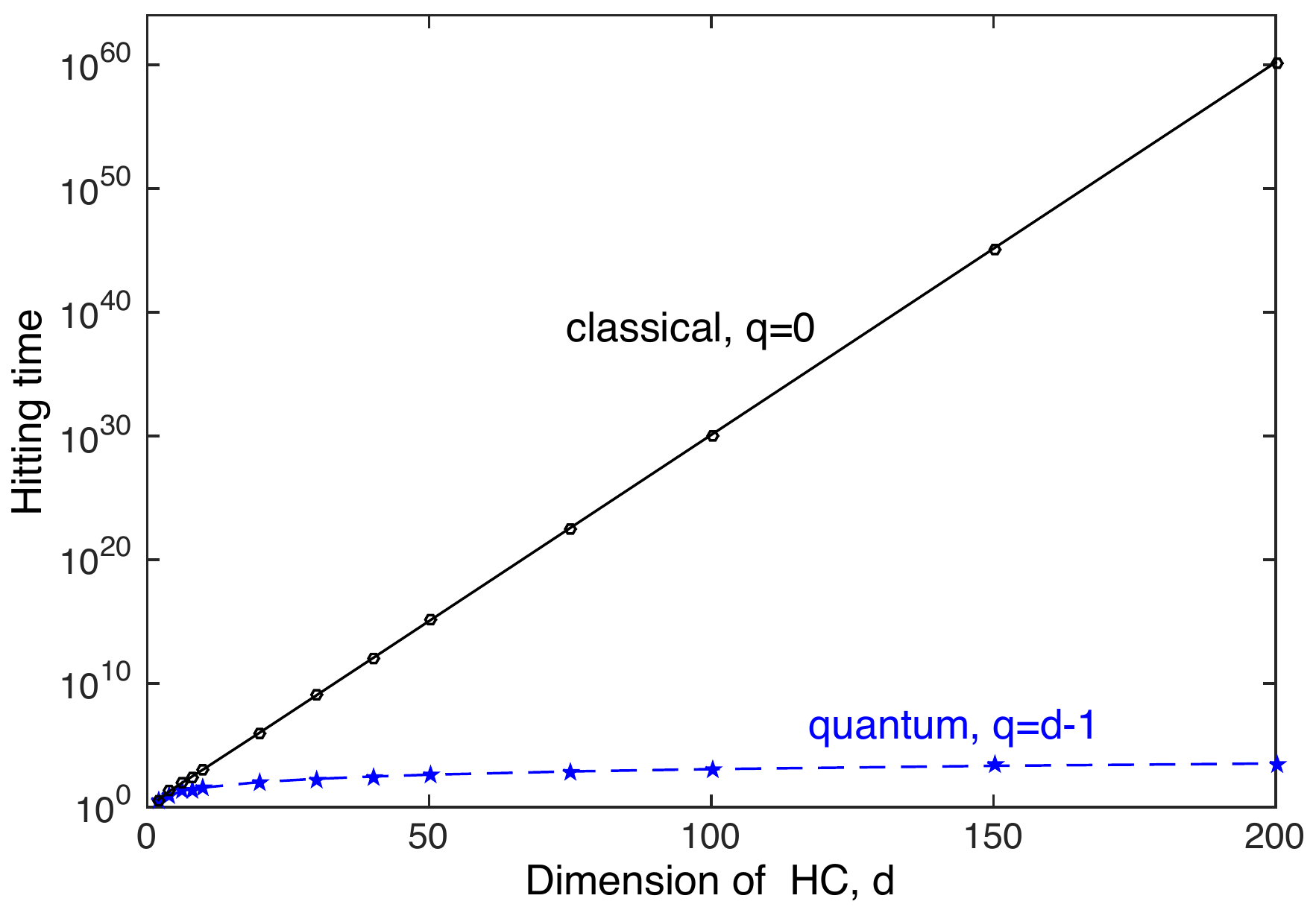}%
	\end{minipage}%
	\begin{minipage}[c]{0.44\linewidth}
		\includegraphics[width=\linewidth]{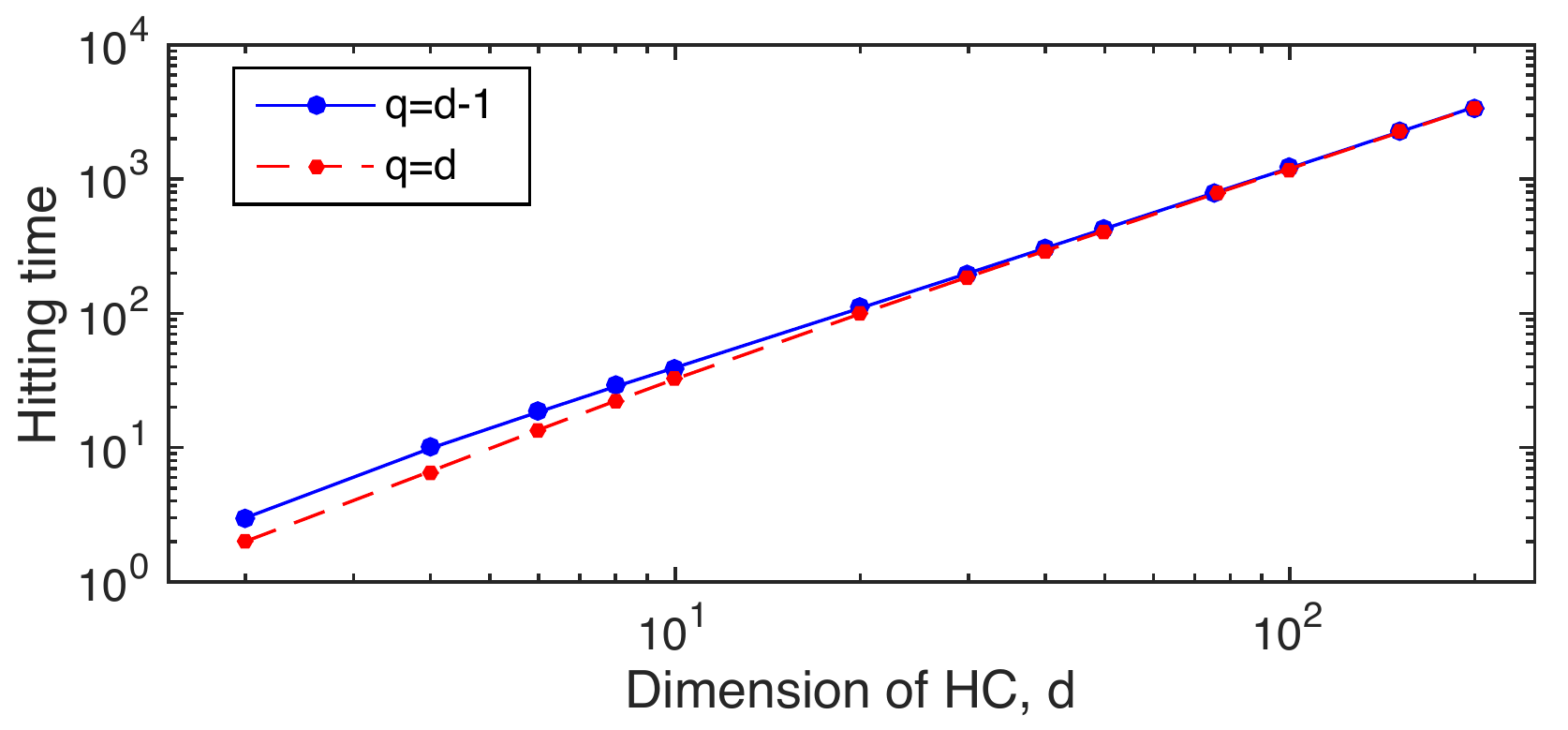}\\
		\includegraphics[width=\linewidth]{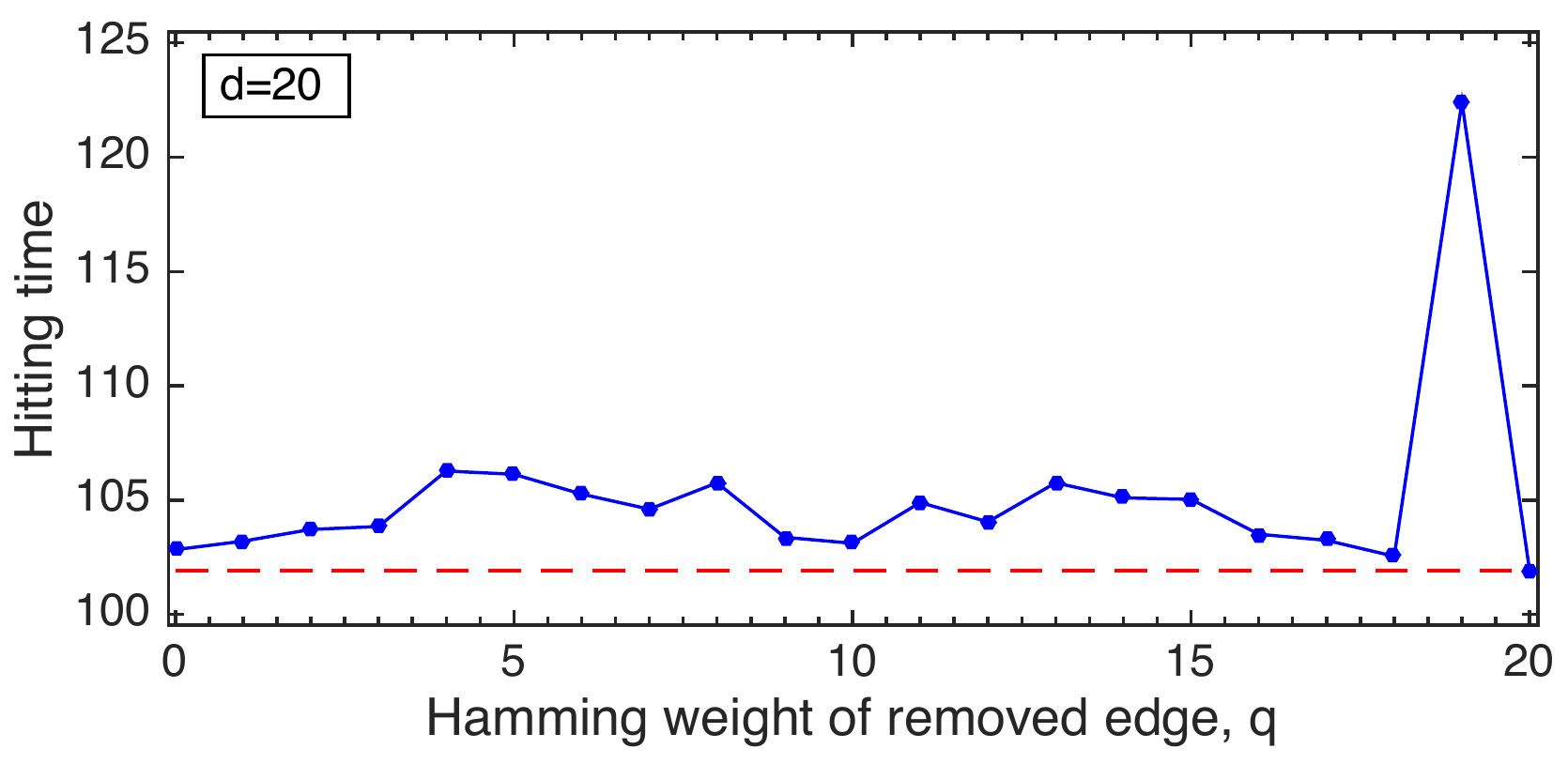}
	\end{minipage}
	\caption{Removed edge: Expected hitting time of a walker on a $d$-dimensional HC with one removed edge between a vertex with Hamming weight $q$ and its neighbor with Hamming weight $q+1$.
	Shown on the left is a comparison of the best classical ($q=0$) and worst quantum ($q=d-1$) walks. The quantum data are from numerical simulations with an error threshold $\epsilon=10^{-4}$.
	Shown on the top right is a comparison of quantum walks with $q=d-1$ (solid blue line) and $q=d$ (the bare HC, dashed red line) on a log-log scale. The data are from simulations with an error threshold $\epsilon=10^{-4}$. The numerical fit to each curve yields a power of $n\simeq 1.5$.
	The bottom right shows a quantum walk with $0\leq q \leq 20$ for fixed $d=20$. The horizontal (dashed red) line represents a quantum walk on the bare HC with $d=20$. The data are from simulations with an error threshold $\epsilon=10^{-12}$.
	}
	\label{fig:removed_edge}
\end{figure*}

Figure~\ref{fig:removed_edge} (left) shows the expected hitting times for both the classical and the quantum walk as a function of the HC dimension~$d$.
For the classical hitting time, we concentrate on the case of $q=0$, i.e., when removing an edge that connects to the starting vertex, which was numerically observed to yield the shortest classical hitting time for all considered values of the HC dimension~$d$.
For the quantum walk we provide the expected hitting times for the case $q=d-1$, which always results in the longest hitting time among the possible choices of $q$ (as in the classical case). It is shown that an exponentially increasing gap opens up between this shortest classical hitting time with $q=0$ and longest quantum hitting time with $q=d-1$. The exponential speed-up of the quantum walk therefore holds also when a single edge is removed from the HC. 

The observation that for the classical walk an edge removed at $q=0$ gives the shortest expected hitting time is also plausible from the following consideration.
Let us assume that despite the introduced asymmetry there is always a uniform distribution of the walker on vertices of the same Hamming weight (this approximation becomes better as $d$ grows). Then the probability to walk right (left), i.e., toward the target state (away from the target state) from the distinguished vertices $x_q$ and $x_{q+1}$ is maximal (minimal) for $q=0$. This necessarily results in a shorter hitting time for $q=0$, compared to the hitting time when edges between higher Hamming weight vertices are deleted.
We observe (data not plotted) a monotonic increase of the classical hitting time as $q$ increases. In comparison with the walk on the regular HC of the same dimension, the perturbed hitting time is shorter for $q=0$ and longer for $q=d-1$. 

The expected hitting times of the quantum walk are shown in Fig.~\ref{fig:removed_edge} (top right) as a function of the HC dimension $d$ on a log-log scale. The case with the longest hitting times for $q=d-1$ is compared to the case of $q=d$, i.e., the regular HC. It is shown that removing an edge from a vertex of Hamming weight $q=d-1$ leads to longer expected hitting times than walking on a unperturbed HC, but that the gap on the log-log scale decreases with $d$. Both appear to be linear, thereby suggesting a scaling of the expected hitting time with a power of the dimension~$d$. A numerical linear fit to both curves yields a power of~$n\simeq 1.5$.

Last, in Fig.~\ref{fig:removed_edge} (bottom right) the expected hitting times of the quantum walk are shown for the case when the HC dimension is fixed to $d=20$ and $q$ is varied over all intermediate $0 \leq q \leq d-1$ values (the case $q=20$ gives the data for the regular HC). It is shown that, in contrast to the classical case, the expected hitting times are not monotonic with $q$, and no apparent order can be observed. Instead, a nontrivial dependence of the expected hitting time on $q$ is obtained, reflecting a convoluted interference pattern of the quantum walk.
We can rule out convergence issues to the degree that numerics were done with error thresholds of $\epsilon=10^{-12}$ and $10^{-14}$.
We conclude that the particular choice of the removed edge does not modify the expected hitting time of the quantum walk to a large extent, with the largest deviation obtained if it is deleted next to the final vertex.

\section{Conclusion} \label{sec:conclusion}

We have extended the family of graphs for which the discrete quantum walk gives rise to an exponential speed-up in terms of the expected hitting time over the corresponding classical walk. In particular, we have considered nonsymmetric and nonlocal structures, realized by hypercubes appended with a single tail and HCs embedded in larger HCs, respectively. In addition, we examined a nonsymmetric perturbation of the HC by removing a single edge. The modified graphs with an added tail or a removed edge are examples of nonregular graphs that exhibit a quantum speed-up.
For all three kinds of perturbations, we studied the walks to the opposite corner on the embedded (or perturbed) HC and we analyzed the expected hitting time of the classical random walks (analytically) and the corresponding quantum walks (numerically). Our numerical analysis confirms that the expected hitting times of the classical walks grow exponentially with the HC dimension $d$, different from their quantum counterparts.

To that end we provided a general mapping procedure that enables the simulation of quantum walks on perturbed HCs of very high dimensions (up to hundreds). With this procedure, the full high-dimensional Hilbert space of the quantum walk, which is exponentially large in the HC dimension $d$, is mapped to an effective subspace, which has a grid structure in the position space and is only polynomially large in $d$. The mapping procedure we supply is general in the sense that it accounts for any number of perturbations on the HC. Specifically, perturbing the HC in $m$ vertices, gives rise to a grid structure of dimension of at most $m+1$. This mapping generalizes the known mapping procedure of the corner-to-corner walk on the bare HC to a line.

In the examples we consider in this paper, the central HC is perturbed in either one or two vertices. This breaks the original symmetry of the bare HC and perturbs the interference pattern of the quantum walk. Nevertheless, it is observed that the exponential speed up of the quantum walk remains. This suggests that quantum walks on graphs that are not fully symmetric can also be exponentially faster than the corresponding classical walks. However, to what extent can this observation hold? Will the exponential gap between the classical and the quantum walk survive also more extensive perturbations? Clearly, the quantum walk is not exponentially faster on \emph{any} graph. One example is the quantum walk on a $q$-dimensional HC embedded in a HC of dimension $d=2q$, which we observe to also scale exponentially. In this context we should note the following: First, much of the original symmetry of the HC remained after the perturbations we applied and second, perturbing a constant number in an exponentially increasing structure is plausibly negligible. The same is also observed in Ref.~\cite{2006a_Krovi_PRA}, where a slight distortion of the HC did not ruin the exponential speed-up of the quantum walk. It would therefore be of interest to check whether the speed-up survives also when the number of perturbed vertices scales with the HC dimension~$d$.

Note, however, that the discussion of the influence of a single perturbation in a graph of exponentially many vertices and edges is not straightforward and that a perturbative picture does not necessarily apply. Although the number of perturbed vertices and edges is exponentially small compared to the size of the graph, the resulting interference patterns in the quantum walk also only require a small change (e.g., changing an interference maximum to a minimum at the final vertex) to yield very different dynamics for hitting the final vertex. An example that we encountered in the present study is the walk on a HC within a larger HC, where just moving the final vertex by two sites to $q=d-2$ results in a suppression by a factor $1/10$ of the walker to reach the final vertex for $d=20$ (Fig.~\ref{fig:hc_in_hc_ptot_and_tau_c}, top) and it decreases further for larger~$d$.

The fact that adding a single vertex or removing a single edge does not qualitatively change the scaling behavior of the expected hitting time with respect to the unperturbed HC and that it results also quantitatively in the same order of magnitude can, to some extent, be rationalized with a picture of the walker coherently taking many possible paths towards the final vertex. According to that picture, a single perturbation affects a certain fraction of these paths by inserting possible detours (in the case of an added vertex) or interrupting these paths (in the case of a removed edge). That picture, which entails that a removed edge close to the final vertex interrupts a larger fraction of paths, is consistent with the observation that the hitting time exhibits a maximum when an edge adjacent to the final vertex is removed (see Fig.~\ref{fig:removed_edge}, bottom right). Another maximum, however, is not observed for a removed edge next to the starting vertex. Furthermore, that picture suggests that with increasing HC dimension $d$ a single perturbation affects a decreasingly smaller fraction of paths. That is consistent with the observation of expected hitting times converging to the same value as $d$ increases in the case of a single removed edge (see Fig.~\ref{fig:removed_edge}, top right). The same observation, however, cannot be made for the case of an added vertex (cf.\ Fig.~\ref{fig:tail_varying_d}, top right), where the hitting times of the different cases appear to remain separated by a constant factor.

We have also observed that the expected hitting time of the quantum walk differs qualitatively from its classical analog, not only in its scaling with the HC dimension, but in other aspects as well. 
In particular, a convoluted nonmonotonic dependence on the Hamming weight of the perturbation $q$ is revealed. This is in sharp contrast to the classical case, where the expected hitting time assumes a monotonic dependence on~$q$.

Another property in which the quantum walk differs significantly from the classical one is the possible existence of dark states: Whereas the classical walk on a connected graph is always guaranteed to end, the quantum walk might become stuck in the graph without ever reaching the target state. For these cases we introduced the conditional hitting time. Numerically, we observed dark states to play a role only in the nonlocal embedding of a HC inside a HC, whenever the dimension $q$ of the embedded HC differs from 1 and from $d-1$, where $d$ is the dimension of the larger HC. In addition, we saw that the probability to become locked inside the subspace of dark states approaches 1 exponentially fast as $\abs{d-q}\rightarrow \tfrac{d}{2}$. By simulating the conditional hitting time for these cases we presented examples that scale polynomially when the embedding is constant ($d=q+1$), whereas the scaling is exponential, as in the classical case, for an embedding that scales with the HC dimension ($d=2q$).


\acknowledgments
We acknowledge support from the Austrian Science Fund (FWF) through the SFB FoQuS Grant No.~F4012, and the Templeton World Charity Foundation Grant No.~TWCF0078/AB46. A.M. and M.T. have contributed equally to this work.


\nocite{apsrev41Control}
\bibliographystyle{apsrev4-1}
\bibliography{bibliography}

\begin{thebibliography}{20}%
\makeatletter
\providecommand \@ifxundefined [1]{%
 \@ifx{#1\undefined}
}%
\providecommand \@ifnum [1]{%
 \ifnum #1\expandafter \@firstoftwo
 \else \expandafter \@secondoftwo
 \fi
}%
\providecommand \@ifx [1]{%
 \ifx #1\expandafter \@firstoftwo
 \else \expandafter \@secondoftwo
 \fi
}%
\providecommand \natexlab [1]{#1}%
\providecommand \enquote  [1]{``#1''}%
\providecommand \bibnamefont  [1]{#1}%
\providecommand \bibfnamefont [1]{#1}%
\providecommand \citenamefont [1]{#1}%
\providecommand \href@noop [0]{\@secondoftwo}%
\providecommand \href [0]{\begingroup \@sanitize@url \@href}%
\providecommand \@href[1]{\@@startlink{#1}\@@href}%
\providecommand \@@href[1]{\endgroup#1\@@endlink}%
\providecommand \@sanitize@url [0]{\catcode `\\12\catcode `\$12\catcode
  `\&12\catcode `\#12\catcode `\^12\catcode `\_12\catcode `\%12\relax}%
\providecommand \@@startlink[1]{}%
\providecommand \@@endlink[0]{}%
\providecommand \url  [0]{\begingroup\@sanitize@url \@url }%
\providecommand \@url [1]{\endgroup\@href {#1}{\urlprefix }}%
\providecommand \urlprefix  [0]{URL }%
\providecommand \Eprint [0]{\href }%
\providecommand \doibase [0]{http://dx.doi.org/}%
\providecommand \selectlanguage [0]{\@gobble}%
\providecommand \bibinfo  [0]{\@secondoftwo}%
\providecommand \bibfield  [0]{\@secondoftwo}%
\providecommand \translation [1]{[#1]}%
\providecommand \BibitemOpen [0]{}%
\providecommand \bibitemStop [0]{}%
\providecommand \bibitemNoStop [0]{.\EOS\space}%
\providecommand \EOS [0]{\spacefactor3000\relax}%
\providecommand \BibitemShut  [1]{\csname bibitem#1\endcsname}%
\let\auto@bib@innerbib\@empty
\bibitem [{\citenamefont {Kempe}(2003)}]{2003_Kempe_review}%
  \BibitemOpen
  \bibfield  {author} {\bibinfo {author} {\bibfnamefont {J.}~\bibnamefont
  {Kempe}},\ }\bibfield  {title} {\enquote {\bibinfo {title} {Quantum random
  walks - {A}n introductory overview},}\ }\href
  {http://dx.doi.org/10.1080/00107151031000110776} {\bibfield  {journal}
  {\bibinfo  {journal} {Contemporary Physics}\ }\textbf {\bibinfo {volume}
  {44}},\ \bibinfo {pages} {307} (\bibinfo {year} {2003})}\BibitemShut
  {NoStop}%
\bibitem [{\citenamefont {Reitzner}\ \emph {et~al.}(2011)\citenamefont
  {Reitzner}, \citenamefont {Nagaj},\ and\ \citenamefont
  {Bu\v{z}ek}}]{2012_Reitzner_review}%
  \BibitemOpen
  \bibfield  {author} {\bibinfo {author} {\bibfnamefont {D.}~\bibnamefont
  {Reitzner}}, \bibinfo {author} {\bibfnamefont {D.}~\bibnamefont {Nagaj}}, \
  and\ \bibinfo {author} {\bibfnamefont {V.}~\bibnamefont {Bu\v{z}ek}},\
  }\bibfield  {title} {\enquote {\bibinfo {title} {Quantum walks},}\ }\href
  {http://www.physics.sk/aps/pub.php?y=2011&pub=aps-11-06} {\bibfield
  {journal} {\bibinfo  {journal} {Acta Physica Slovaca}\ }\textbf {\bibinfo
  {volume} {61}},\ \bibinfo {pages} {603--725} (\bibinfo {year}
  {2011})}\BibitemShut {NoStop}%
\bibitem [{\citenamefont {Kempe}(2005)}]{2003_Kempe}%
  \BibitemOpen
  \bibfield  {author} {\bibinfo {author} {\bibfnamefont {J.}~\bibnamefont
  {Kempe}},\ }\bibfield  {title} {\enquote {\bibinfo {title} {Discrete quantum
  random walks hit exponentially faster},}\ }\href
  {http://dx.doi.org/10.1007/s00440-004-0423-2} {\bibfield  {journal} {\bibinfo
   {journal} {Probab. Theory Relat. Fields}\ }\textbf {\bibinfo {volume}
  {133}},\ \bibinfo {pages} {215--235} (\bibinfo {year} {2005})}\BibitemShut
  {NoStop}%
\bibitem [{\citenamefont {Childs}\ \emph {et~al.}(2003)\citenamefont {Childs},
  \citenamefont {Cleve}, \citenamefont {Deotto}, \citenamefont {Farhi},
  \citenamefont {Gutmann},\ and\ \citenamefont {Spielman}}]{2003_Childs}%
  \BibitemOpen
  \bibfield  {author} {\bibinfo {author} {\bibfnamefont {A.~M.}\ \bibnamefont
  {Childs}}, \bibinfo {author} {\bibfnamefont {R.}~\bibnamefont {Cleve}},
  \bibinfo {author} {\bibfnamefont {E.}~\bibnamefont {Deotto}}, \bibinfo
  {author} {\bibfnamefont {E.}~\bibnamefont {Farhi}}, \bibinfo {author}
  {\bibfnamefont {S.}~\bibnamefont {Gutmann}}, \ and\ \bibinfo {author}
  {\bibfnamefont {D.~A.}\ \bibnamefont {Spielman}},\ }\bibfield  {title}
  {\enquote {\bibinfo {title} {Exponential algorithmic speedup by a quantum
  walk},}\ }in\ \href {\doibase 10.1145/780542.780552} {\emph {\bibinfo
  {booktitle} {Proceedings of the 35th {A}nnual ACM {S}ymposium on {T}heory of
  {C}omputing}}}\ (\bibinfo  {publisher} {ACM},\ \bibinfo {address} {New
  York},\ \bibinfo {year} {2003})\ pp.\ \bibinfo {pages} {59--68}\BibitemShut
  {NoStop}%
\bibitem [{\citenamefont {Nayak}\ and\ \citenamefont
  {Vishwanath}(2000)}]{2000_Nayak_Arx}%
  \BibitemOpen
  \bibfield  {author} {\bibinfo {author} {\bibfnamefont {A.}~\bibnamefont
  {Nayak}}\ and\ \bibinfo {author} {\bibfnamefont {A.}~\bibnamefont
  {Vishwanath}},\ }\bibfield  {title} {\enquote {\bibinfo {title} {Quantum walk
  on the line},}\ }\href {http://arxiv.org/abs/quant-ph/0010117} {\bibfield
  {journal} {\bibinfo  {journal} {Preprint arXiv:quant-ph/0010117}\ } (\bibinfo
  {year} {2000})}\BibitemShut {NoStop}%
\bibitem [{\citenamefont {Ambainis}\ \emph {et~al.}(2001)\citenamefont
  {Ambainis}, \citenamefont {Bach}, \citenamefont {Nayak}, \citenamefont
  {Vishwanath},\ and\ \citenamefont {Watrous}}]{2001_Ambainis_Proceedings}%
  \BibitemOpen
  \bibfield  {author} {\bibinfo {author} {\bibfnamefont {A.}~\bibnamefont
  {Ambainis}}, \bibinfo {author} {\bibfnamefont {E.}~\bibnamefont {Bach}},
  \bibinfo {author} {\bibfnamefont {A.}~\bibnamefont {Nayak}}, \bibinfo
  {author} {\bibfnamefont {A.}~\bibnamefont {Vishwanath}}, \ and\ \bibinfo
  {author} {\bibfnamefont {J.}~\bibnamefont {Watrous}},\ }\bibfield  {title}
  {\enquote {\bibinfo {title} {One-dimensional quantum walks},}\ }in\ \href
  {http://doi.acm.org/10.1145/380752.380757} {\emph {\bibinfo {booktitle}
  {Proceedings of the 33rd {A}nnual ACM {S}ymposium on {T}heory of
  {C}omputing}}}\ (\bibinfo  {publisher} {ACM},\ \bibinfo {address} {New
  York},\ \bibinfo {year} {2001})\ pp.\ \bibinfo {pages} {37--49}\BibitemShut
  {NoStop}%
\bibitem [{\citenamefont {Aharonov}\ \emph {et~al.}(2001)\citenamefont
  {Aharonov}, \citenamefont {Ambainis}, \citenamefont {Kempe},\ and\
  \citenamefont {Vazirani}}]{2001_Aharonov_Proceedings}%
  \BibitemOpen
  \bibfield  {author} {\bibinfo {author} {\bibfnamefont {D.}~\bibnamefont
  {Aharonov}}, \bibinfo {author} {\bibfnamefont {A.}~\bibnamefont {Ambainis}},
  \bibinfo {author} {\bibfnamefont {J.}~\bibnamefont {Kempe}}, \ and\ \bibinfo
  {author} {\bibfnamefont {U.}~\bibnamefont {Vazirani}},\ }\bibfield  {title}
  {\enquote {\bibinfo {title} {Quantum walks on graphs},}\ }in\ \href
  {http://doi.acm.org/10.1145/380752.380758} {\emph {\bibinfo {booktitle}
  {Proceedings of the 33rd {A}nnual ACM {S}ymposium on {T}heory of
  {C}omputing}}}\ (\bibinfo  {publisher} {ACM},\ \bibinfo {address} {New
  York},\ \bibinfo {year} {2001})\ pp.\ \bibinfo {pages} {50--59}\BibitemShut
  {NoStop}%
\bibitem [{\citenamefont {Krovi}\ and\ \citenamefont
  {Brun}(2006{\natexlab{a}})}]{2006a_Krovi_PRA}%
  \BibitemOpen
  \bibfield  {author} {\bibinfo {author} {\bibfnamefont {H.}~\bibnamefont
  {Krovi}}\ and\ \bibinfo {author} {\bibfnamefont {T.~A.}\ \bibnamefont
  {Brun}},\ }\bibfield  {title} {\enquote {\bibinfo {title} {Hitting time for
  quantum walks on the hypercube},}\ }\href
  {http://link.aps.org/doi/10.1103/PhysRevA.73.032341} {\bibfield  {journal}
  {\bibinfo  {journal} {Phys. Rev. A}\ }\textbf {\bibinfo {volume} {73}},\
  \bibinfo {pages} {032341} (\bibinfo {year} {2006}{\natexlab{a}})}\BibitemShut
  {NoStop}%
\bibitem [{\citenamefont {Makmal}\ \emph {et~al.}(2014)\citenamefont {Makmal},
  \citenamefont {Zhu}, \citenamefont {Manzano}, \citenamefont {Tiersch},\ and\
  \citenamefont {Briegel}}]{2014_HCI_PRA}%
  \BibitemOpen
  \bibfield  {author} {\bibinfo {author} {\bibfnamefont {A.}~\bibnamefont
  {Makmal}}, \bibinfo {author} {\bibfnamefont {M.}~\bibnamefont {Zhu}},
  \bibinfo {author} {\bibfnamefont {D.}~\bibnamefont {Manzano}}, \bibinfo
  {author} {\bibfnamefont {M.}~\bibnamefont {Tiersch}}, \ and\ \bibinfo
  {author} {\bibfnamefont {H.~J.}\ \bibnamefont {Briegel}},\ }\bibfield
  {title} {\enquote {\bibinfo {title} {Quantum walks on embedded hypercubes},}\
  }\href {http://link.aps.org/doi/10.1103/PhysRevA.90.022314} {\bibfield
  {journal} {\bibinfo  {journal} {Phys. Rev. A}\ }\textbf {\bibinfo {volume}
  {90}} (\bibinfo {year} {2014})}\BibitemShut {NoStop}%
\bibitem [{\citenamefont {Krovi}\ and\ \citenamefont
  {Brun}(2006{\natexlab{b}})}]{2006b_Krovi_PRA}%
  \BibitemOpen
  \bibfield  {author} {\bibinfo {author} {\bibfnamefont {H.}~\bibnamefont
  {Krovi}}\ and\ \bibinfo {author} {\bibfnamefont {T.~A.}\ \bibnamefont
  {Brun}},\ }\bibfield  {title} {\enquote {\bibinfo {title} {Quantum walks with
  infinite hitting times},}\ }\href {\doibase 10.1103/PhysRevA.74.042334}
  {\bibfield  {journal} {\bibinfo  {journal} {Phys. Rev. A}\ }\textbf {\bibinfo
  {volume} {74}},\ \bibinfo {pages} {042334} (\bibinfo {year}
  {2006}{\natexlab{b}})}\BibitemShut {NoStop}%
\bibitem [{\citenamefont {Kac}(1947)}]{Kac}%
  \BibitemOpen
  \bibfield  {author} {\bibinfo {author} {\bibfnamefont {M.}~\bibnamefont
  {Kac}},\ }\bibfield  {title} {\enquote {\bibinfo {title} {Random walk and the
  theory of {B}rownian motion},}\ }\href {\doibase 10.2307/2304386} {\bibfield
  {journal} {\bibinfo  {journal} {Am.\ Math.\ Mon.}\ }\textbf {\bibinfo
  {volume} {54}},\ \bibinfo {pages} {369--391} (\bibinfo {year}
  {1947})}\BibitemShut {NoStop}%
\bibitem [{\citenamefont {Moore}\ and\ \citenamefont
  {Russell}(2002)}]{2002_Moore_inBook}%
  \BibitemOpen
  \bibfield  {author} {\bibinfo {author} {\bibfnamefont {C.}~\bibnamefont
  {Moore}}\ and\ \bibinfo {author} {\bibfnamefont {A.}~\bibnamefont
  {Russell}},\ }\bibfield  {title} {\enquote {\bibinfo {title} {Quantum walks
  on the hypercube},}\ }in\ \href {http://dx.doi.org/10.1007/3-540-45726-7_14}
  {\emph {\bibinfo {booktitle} {Randomization and {A}pproximation {T}echniques
  in {C}omputer {S}cience}}},\ Vol.\ \bibinfo {volume} {2483},\ \bibinfo
  {editor} {edited by\ \bibinfo {editor} {\bibfnamefont {J.~D.~P.}\
  \bibnamefont {Rolim}}\ and\ \bibinfo {editor} {\bibfnamefont
  {S.}~\bibnamefont {Vadhan}}}\ (\bibinfo  {publisher} {Springer, Berlin},\
  \bibinfo {year} {2002})\ pp.\ \bibinfo {pages} {164--178}\BibitemShut
  {NoStop}%
\bibitem [{\citenamefont {Grinstead}\ and\ \citenamefont
  {Snell}(1997)}]{GrinsteadSnell1997}%
  \BibitemOpen
  \bibfield  {author} {\bibinfo {author} {\bibfnamefont {C.~M.}\ \bibnamefont
  {Grinstead}}\ and\ \bibinfo {author} {\bibfnamefont {J.~L.}\ \bibnamefont
  {Snell}},\ }\href@noop {} {\emph {\bibinfo {title} {Introduction to
  {P}robability}}},\ 2nd ed\ (\bibinfo  {publisher} {AMS},\ \bibinfo {address}
  {Providence},\ \bibinfo {year} {1997})\BibitemShut {NoStop}%
\bibitem [{\citenamefont {Novo}\ \emph {et~al.}(2015)\citenamefont {Novo},
  \citenamefont {Chakraborty}, \citenamefont {Mohseni}, \citenamefont {Neven},\
  and\ \citenamefont {Omar}}]{2014_Novo_arXiv}%
  \BibitemOpen
  \bibfield  {author} {\bibinfo {author} {\bibfnamefont {L.}~\bibnamefont
  {Novo}}, \bibinfo {author} {\bibfnamefont {S.}~\bibnamefont {Chakraborty}},
  \bibinfo {author} {\bibfnamefont {M.}~\bibnamefont {Mohseni}}, \bibinfo
  {author} {\bibfnamefont {H.}~\bibnamefont {Neven}}, \ and\ \bibinfo {author}
  {\bibfnamefont {Y.}~\bibnamefont {Omar}},\ }\bibfield  {title} {\enquote
  {\bibinfo {title} {Systematic dimensionality reduction for quantum walks:
  Optimal spatial search and transport on non-regular graphs},}\ }\href
  {\doibase 10.1038/srep13304} {\bibfield  {journal} {\bibinfo  {journal}
  {Sci.\ Rep.}\ }\textbf {\bibinfo {volume} {5}},\ \bibinfo {pages} {13304}
  (\bibinfo {year} {2015})}\BibitemShut {NoStop}%
\bibitem [{\citenamefont {Voit}(1996)}]{Voit}%
  \BibitemOpen
  \bibfield  {author} {\bibinfo {author} {\bibfnamefont {M.}~\bibnamefont
  {Voit}},\ }\bibfield  {title} {\enquote {\bibinfo {title} {Asymptotic
  distribution for the {E}hrenfest urn and related random walks},}\ }\href
  {\doibase 10.2307/3215058} {\bibfield  {journal} {\bibinfo  {journal} {J.
  Appl.\ Prob.}\ }\textbf {\bibinfo {volume} {33}},\ \bibinfo {pages}
  {340--356} (\bibinfo {year} {1996})}\BibitemShut {NoStop}%
\bibitem [{\citenamefont {Donnelly}\ \emph {et~al.}(1994)\citenamefont
  {Donnelly}, \citenamefont {Lloyd},\ and\ \citenamefont {Sudbury}}]{Donnelly}%
  \BibitemOpen
  \bibfield  {author} {\bibinfo {author} {\bibfnamefont {P.}~\bibnamefont
  {Donnelly}}, \bibinfo {author} {\bibfnamefont {P.}~\bibnamefont {Lloyd}}, \
  and\ \bibinfo {author} {\bibfnamefont {A.}~\bibnamefont {Sudbury}},\
  }\bibfield  {title} {\enquote {\bibinfo {title} {Approach to stationarity of
  the {B}ernoulli-{L}aplace diffusion model},}\ }\href
  {http://www.jstor.org/stable/1427817} {\bibfield  {journal} {\bibinfo
  {journal} {Adv.\ Appl.\ Prob.}\ }\textbf {\bibinfo {volume} {26}},\ \bibinfo
  {pages} {715--727} (\bibinfo {year} {1994})}\BibitemShut {NoStop}%
\bibitem [{\citenamefont {Hutton}(1980)}]{Hutton}%
  \BibitemOpen
  \bibfield  {author} {\bibinfo {author} {\bibfnamefont {J.}~\bibnamefont
  {Hutton}},\ }\bibfield  {title} {\enquote {\bibinfo {title} {The recurrence
  and transience of two-dimensional linear birth and death processes},}\ }\href
  {\doibase 10.2307/1426423} {\bibfield  {journal} {\bibinfo  {journal} {Adv.\
  Appl.\ Prob.}\ }\textbf {\bibinfo {volume} {12}},\ \bibinfo {pages}
  {615--639} (\bibinfo {year} {1980})}\BibitemShut {NoStop}%
\bibitem [{\citenamefont {Moisy}(2014)}]{EZYFIT_MATLAB}%
  \BibitemOpen
  \bibfield  {author} {\bibinfo {author} {\bibfnamefont {F.}~\bibnamefont
  {Moisy}},\ }\href {http://www.fast.u-psud.fr/ezyfit/} {\bibfield  {journal}
  {\bibinfo  {journal} {\textsc{EzyFit {M}atlab} toolbox, Version 2.4,
  Available at http://www.fast.u-psud.fr/ezyfit/}\ } (\bibinfo {year}
  {2014})}\BibitemShut {NoStop}%
\bibitem [{\citenamefont {Varbanov}\ \emph {et~al.}(2008)\citenamefont
  {Varbanov}, \citenamefont {Krovi},\ and\ \citenamefont
  {Brun}}]{2008_Krovi_PRA}%
  \BibitemOpen
  \bibfield  {author} {\bibinfo {author} {\bibfnamefont {M.}~\bibnamefont
  {Varbanov}}, \bibinfo {author} {\bibfnamefont {H.}~\bibnamefont {Krovi}}, \
  and\ \bibinfo {author} {\bibfnamefont {T.~A.}\ \bibnamefont {Brun}},\
  }\bibfield  {title} {\enquote {\bibinfo {title} {Hitting time for the
  continuous quantum walk},}\ }\href {\doibase 10.1103/PhysRevA.78.022324}
  {\bibfield  {journal} {\bibinfo  {journal} {Phys. Rev. A}\ }\textbf {\bibinfo
  {volume} {78}},\ \bibinfo {pages} {022324} (\bibinfo {year}
  {2008})}\BibitemShut {NoStop}%
\bibitem [{\citenamefont {Fleischhauer}\ \emph {et~al.}(2005)\citenamefont
  {Fleischhauer}, \citenamefont {Imamoglu},\ and\ \citenamefont
  {Marangos}}]{Fleischhauer}%
  \BibitemOpen
  \bibfield  {author} {\bibinfo {author} {\bibfnamefont {M.}~\bibnamefont
  {Fleischhauer}}, \bibinfo {author} {\bibfnamefont {A.}~\bibnamefont
  {Imamoglu}}, \ and\ \bibinfo {author} {\bibfnamefont {J.~P.}\ \bibnamefont
  {Marangos}},\ }\bibfield  {title} {\enquote {\bibinfo {title}
  {Electromagnetically induced transparency: Optics in coherent media},}\
  }\href {\doibase 10.1103/RevModPhys.77.633} {\bibfield  {journal} {\bibinfo
  {journal} {Rev. Mod. Phys.}\ }\textbf {\bibinfo {volume} {77}},\ \bibinfo
  {pages} {633--673} (\bibinfo {year} {2005})}\BibitemShut {NoStop}%
\end{thebibliography}%


\end{document}